\documentclass[12pt,preprint]{aastex}
\usepackage{xspace}
\usepackage{graphicx}
\usepackage{natbib}
\graphicspath{{.}}
\usepackage{epsfig}
\usepackage{epstopdf}
\newcommand{\nraoblurb}{The National Radio Astronomy Observatory is
a facility of the National Science Foundation operated under cooperative
agreement by Associated Universities, Inc.}
\newcommand{\hide}[1]{}

\newcommand{\gl}{\ensuremath{\ell}\xspace}
\newcommand{\gb}{\ensuremath{{\it b}}\xspace}

\newcommand{\lb}{\ensuremath{(\gl,\gb)}\xspace}
\newcommand{\lv}{\ensuremath{(\gl,v)}\xspace}

\newcommand{\kms}{\ensuremath{\,{\rm km\,s^{-1}}}\xspace}

\newcommand{\microns}{\ensuremath{\,\mu{\rm m}}\xspace}

\newcommand{\degree}{\ensuremath{\,^\circ}\xspace}

\newcommand{\hi}{{\rm H\,{\footnotesize I}}\xspace}
\newcommand{\hii}{{\rm H\,{\footnotesize II}}\xspace}

\newcommand{\co} {\ensuremath{^{\rm 12}{\rm CO}}\xspace}
\newcommand{\cor}{\ensuremath{^{\rm 13}{\rm CO}}\xspace}

\slugcomment{After referee report --- lda, \today}
\shorttitle{WISE \hii\ Region Catalog}
\shortauthors{Anderson et al.}

\begin{document}

\title{The WISE Catalog of Galactic H\,{\small \bf II} Regions}

\author{L.~D.~Anderson\altaffilmark{1, 2},
  T.~M.~Bania\altaffilmark{3}, Dana~S.~Balser\altaffilmark{4},
  V.~Cunningham\altaffilmark{1}, T.~V.~Wenger\altaffilmark{4,5},
  B.~M.~Johnstone\altaffilmark{1}, W.~P.~Armentrout\altaffilmark{1}}

\altaffiltext{1}{Department of Physics and Astronomy, West Virginia
  University, Morgantown, West Virginia, USA}
\altaffiltext{2}{National Radio Astronomy Observatory, PO Box 2, Green
  Bank, WV 24944, USA}
\altaffiltext{3}{Institute for Astrophysical Research, Department of
  Astronomy, Boston University, 725 Commonwealth Avenue, Boston, MA
  02215, USA}
\altaffiltext{4}{National Radio Astronomy Observatory, 520 Edgemont
  Road, Charlottesville VA, 22903, USA}
\altaffiltext{5}{Department of Astronomy, University of Virginia,
  P.O. Box 3813, Charlottesville, VA 22904, USA}

%%%%%%%%%%%%%%%%%%%%%%%%%%%%%%%%%%%%%%%%%%%%%%%%%%
\begin{abstract}
Using data from the all-sky Wide-Field Infrared Survey Explorer (WISE)
satellite, we made a catalog of over 8000 Galactic \hii\ regions and
\hii\ region candidates by searching for their characteristic
mid-infrared (MIR) morphology.  WISE has sufficient sensitivity to
detect the MIR emission from \hii\ regions located anywhere in the
Galactic disk.  We believe this is the most complete catalog yet of
regions forming massive stars in the Milky Way.  Of the $\sim 8000$ cataloged sources, 
$\sim1500$ have measured radio recombination line (RRL) or
H$\alpha$ emission, and are thus known to be \hii\ regions.  This
sample improves on previous efforts by resolving \hii\ region
complexes into multiple sources and by removing duplicate entries.
There are $\sim2500$ candidate \hii\ regions in the catalog that are spatially coincident
with radio continuum emission.  Our group's previous RRL studies show
that $\sim95\%$ of such targets are \hii\ regions.  We find that
$\sim500$ of these candidates are also positionally associated with
known \hii\ region complexes, so the probability of their being {\it
  bona fide} \hii\ regions is even higher.  At the sensitivity limits
of existing surveys, $\sim4000$ catalog sources show no radio continuum
emission.  Using data from the literature, we find distances for
$\sim1500$ catalog sources, and molecular velocities for $\sim1500$
\hii\ region candidates.

\end{abstract}


%%%%%%%%%%%%%%%%%%%%%%%%%%%%%%%%%%%%%%%%%%%%%%%%%%
\section{INTRODUCTION\label{sec:intro}}

\hii\ regions are the zones of ionized gas surrounding young massive
stars.  The stars capable of producing the ultra-violet photons
necessary to ionize their surrounding medium have spectral types of $\sim$B0
or earlier.  Such stars only live $\sim10$\,Myr and thus \hii\ regions
are zero-age objects compared to the age of the Milky Way: they trace
star formation at the present epoch.  \hii\ regions are the brightest
objects in the Galaxy at infrared (IR) and radio wavelengths and can
be detected across the entire Galactic disk.  Unlike other tracers of
Galactic star formation, the identification of an \hii\ region
unambiguously locates massive star formation.  They are the
archetypical tracers of spiral arms and have been instrumental in
creating a better understanding of the structure of our Galaxy.  Their
chemical abundances represent Galactic abundances today, and reveal
the effects of billions of years of Galactic chemical evolution.  They
are the main contributors to the ionized photons in a galaxy, the
emission from which is used to determine extragalactic and Galactic
star formation rates.  In short, Galactic \hii\ regions are extremely
important objects for learning about a number of problems in astrophysics,
including star formation, Galactic structure, and Galactic evolution.

Despite their importance, the census of Galactic \hii\ regions is
severely incomplete, as evidenced by the recent Green Bank Telescope
\hii\ Region Discovery Survey \citep[GBT HRDS;][]{bania10,
  anderson11}.  The GBT HRDS measured the radio recombination line
(RRL) and radio continuum emission from 448 previously unknown
Galactic \hii\ regions.  Over the survey zone, the GBT HRDS doubled
the census of known \hii\ regions with measured RRL emission.  The
average on-source integration time in the GBT HRDS was only $\sim10$
minutes; \citet{anderson11} found hundreds more candidate
\hii\ regions that would have required longer integrations.  This
hints at a larger population of Galactic \hii\ regions about which
nothing is known.

As the GBT HRDS demonstrated, Galactic \hii\ regions can be easily and
reliably identified from mid-infrared (MIR) data.  If the resolution
of the MIR data is sufficient, all \hii\ regions have essentially the
same mid-infrared (MIR) morphology: their $\sim10\,\micron$ emission
surrounds their $\sim20\,\micron$ emission and the latter is
coincident with the ionized gas traced by radio continuum
emission \citep[see][]{anderson11}.  
This characteristic morphology allows one to identify \hii\ region
candidates in MIR images. Radio continuum and RRL observations can
then confirm that these targets are \hii\ regions.  The identification
of young \hii\ regions in IR data, where the emission is from heated
dust, has motivated much of the \hii\ region research over the past 25
years \citep[e.g.][]{wc89a, kurtz94}.

In the GBT HRDS, \citet{anderson11} used data from the {\it Spitzer}
legacy 24\,\micron\ MIPSGAL survey \citep{carey09} to identify
targets.  The {\it Spitzer} legacy surveys were generally limited to
within $1\degree$ of the Galactic mid-plane, and $|\ell| \le
65\degree$.  Most regions of massive star formation are within this
Galactic zone, but a complete sample of Galactic regions forming
massive stars requires coverage outside the zone surveyed by {\it
  Spitzer.}

Data from the all-sky Wide-Field Infrared Survey Explorer (WISE) can
also be used to identify \hii\ regions.  WISE covers the entire sky in
four photometric bands: 3.4\,\micron, 4.6\,\micron, 12\,\micron, and
22\,\micron\ at angular resolutions of 6.1\arcsec, 6.4\arcsec,
  6.5\arcsec, and 12\arcsec, respectively. \hii\ regions appear
visually similar in the WISE 12\,\micron\ and 22\,\micron\ bands
compared with the {\it Spitzer} IRAC \citep{fazio04} 8.0\,\micron\ and
MIPS \citep{rieke04} 24\,\micron\ bands, respectively
\citep[see][]{anderson12a}.  For \hii\ regions, the 8.0\,\micron\ and
12\,\micron\ emission are both largely due to PAH molecules, which
fluoresce in ultra-violet radiation fields.  The IRAC
8.0\,\micron\ band contains strong PAH emission at 7.7\,\micron\ and
8.6\,\micron, whereas the WISE 12\,\micron\ band contains PAH emission
at 11.2\,\micron\ and 12.7\,\micron\ \citep[see][for a
  review]{tielens08}.  The MIPS 24\,\micron\ band actually has a very
similar bandpass compared with the WISE band.  For \hii\ regions this
band traces stochastically heated small dust grains \citep{watson08,
  watson09, deharveng10, anderson12b}.  The WISE 22\,\micron\ band
resolution and sensitivity ($12\arcsec$ and 6\,mJy) are comparable to
those of MIPSGAL ($6\arcsec$ and 1.3\,mJy).

%Here we report an all-sky catalog of WISE-identified \hii\ regions and
%\hii\ region candidates.

%\clearpage
\section{CREATING THE WISE CATALOG OF GALACTIC H\,{\small \bf II} REGIONS\label{sec:sample}}
\begin{figure}[!ht]
\begin{centering}
\includegraphics[width=4.5in]{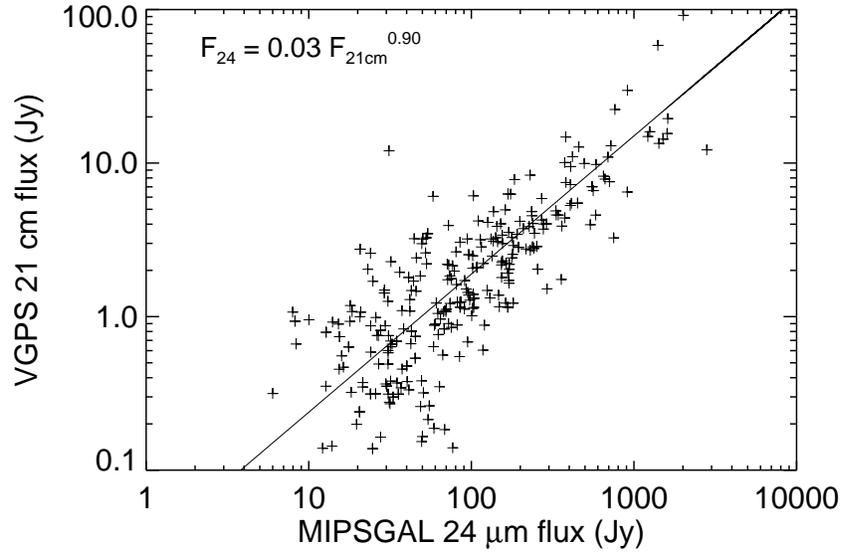}
\caption{Correlation between 24\,\micron\ and 21\,cm fluxes for
  Galactic \hii\ regions.  The data points are from a sample of 301
  \hii\ regions from $15\degree \le \ell \le 55\degree$, $|b| \le
  1\degree$ \citep{anderson_thesis}.  The radio and MIR fluxes are
  highly correlated.  The larger scatter at lower fluxes is likely
  due to photometric errors and uncertainties in the background
  estimation.}
\label{fig:ir_corr}
\end{centering}
\end{figure}

\begin{figure}[!ht]
\begin{centering}
\includegraphics[width=4.5in]{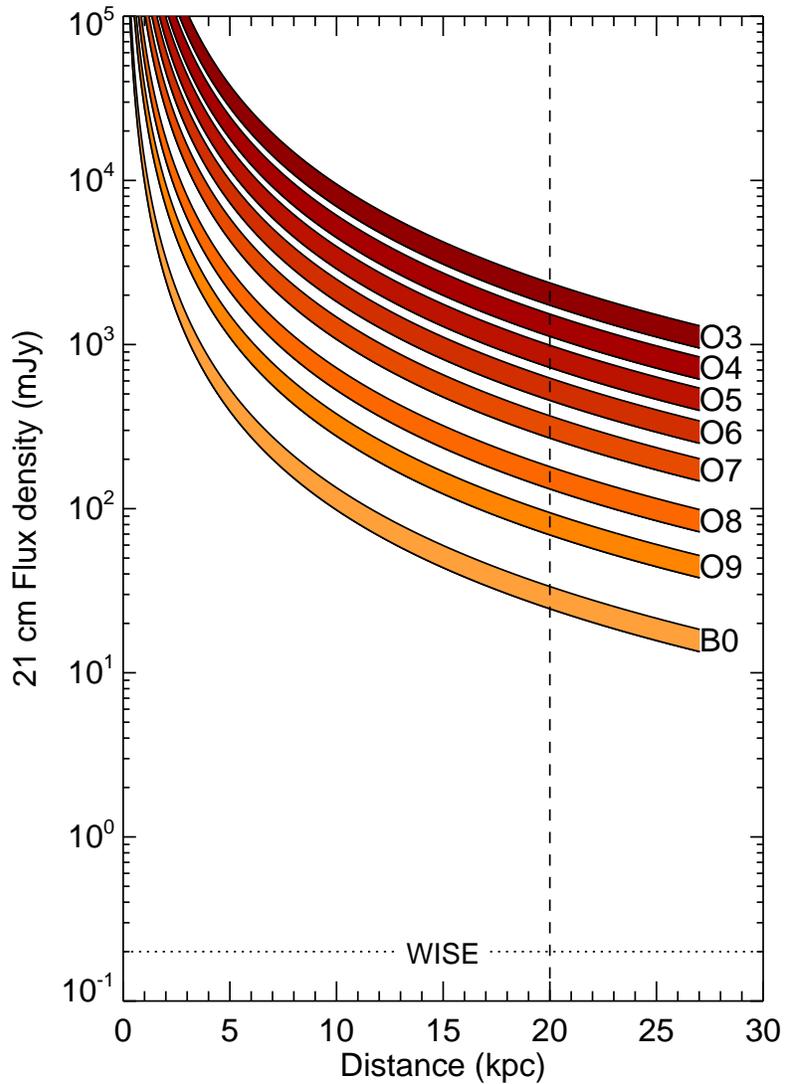}
\caption{Model 21\,cm flux densities for \hii\ regions ionized by
  single stars of spectral types O3 to B0 as a function of
    distance from the Sun.  For the
calculation of the ionizing flux we used \citet{sternberg03} and for
the conversion from ionizing flux to 21\,cm luminosity we used the
relation given in \citet{rubin68}.  The width of the curves reflects a range
  of nebular electron temperatures from 5,000\,K to 10,000\,K.
  The vertical dotted line marks the most distant \hii\ region
    currently known, $\sim20$\,kpc.  The horizontal dotted line is
  the expected radio flux for an \hii\ region at the sensitivity limit
  of the 22\,\micron\ WISE data, $\sim0.2$\,mJy.  WISE has the sensitivity
  to detect all Galactic \hii\ regions.}
\label{fig:flux_v_distance}
\end{centering}
\end{figure}

WISE can in principle detect the MIR emission from all Galactic
\hii\ regions.  \citet{anderson_thesis} measured the integrated
MIPSGAL flux and 21\,cm VLA Galactic Plane Survey
\citep[VGPS;][]{stil06} continuum emission from a large sample of 301
first quadrant Galactic \hii\ regions.  Using these data,
Figure~\ref{fig:ir_corr} shows that the emission at 21\,cm wavelength is $\sim
30$ times less than that at 24\,\micron.  \citet{anderson12a} found
that the WISE 22\,\micron\ flux of \hii\ regions is the same as the
MIPSGAL 24\,\micron\ flux.  The sensitivity of WISE, 6\,mJy at
22\,\micron, is therefore able to detect \hii\ regions with integrated
21\,cm fluxes of $\sim0.2$\,mJy.  We show in
Figure~\ref{fig:flux_v_distance} the expected flux for \hii\ regions
ionized by single stars of various spectral types.  For the
calculation of the ionizing flux we used \citet{sternberg03} and for
the conversion from ionizing flux to 21\,cm luminosity we used the
relation given in \citet{rubin68}.  The expected 21\,cm flux for an
\hii\ region at the sensitivity limit of WISE, using the 30:1 ratio
found for the ratio of the MIPSGAL to VGPS fluxes, is well below that
required to detect the MIR emission from all Galactic \hii\ regions.

This holds true even when extinction is factored into the
calculations.  \citet{flaherty07} find that for {\it Spitzer} the
extinction at 24\,\micron\ is about half that at
2.16\,\micron\ ($K_{S}$ band).  Because the 22\,\micron\ WISE filter
is similar to the 24\,\micron\ {\it Spitzer} filter, the
22\,\micron\ WISE band will share essentially the same value.  The
extinction in the $2.16\,\micron\ K_S$ band is nominally about a tenth
that of visual \citep{rieke85}.  Therefore, $A_{24}/A_V \simeq 0.05$
and even 50 magnitudes of visual extinction would only result in 2.5
magnitudes of 22\,\micron\ extinction.  Two and a half magnitudes of
extinction corresponds to an intensity decrease of a factor of 10.
Even after applying this factor, the WISE sensitivity in
Figure~\ref{fig:flux_v_distance} is well-below the flux from an
\hii\ region ionized by a single B0 star anywhere in the Galaxy.
Furthermore, due to the warp of the Galactic disk, the most distant
\hii\ regions are generally found above and below the Galactic
mid-plane, where the line-of-sight extinction is lower.  
In sum, the MIR line-of-sight extinction is
sufficiently low so that we can detect extremely distant \hii\ regions 
with WISE.
%This fact allows us to detect
%extremely distant \hii\ regions with WISE, essentially free from
%concerns about extinction.
%One caveat here is that the given WISE sensitivity is given for point
%sources.  Large diffuse \hii\ regions may have very low surface
%brightness values.  Since these

We use visual and automatic searches of WISE data to identify
\hii\ regions from their MIR emission morphology. We visually search
WISE 12\,\micron\ and 22\,\micron\ images spanning the entire Galactic
plane within 8 degrees of the nominal mid-plane, $|b| \le 8\arcdeg$.
We create WISE 12\,\micron\ and 22\,\micron\ mosaics using the Montage
software\footnote{http://montage.ipac.caltech.edu/}.  The WISE public
image tiles are $1\fdg564\times 1\fdg564$.  We combine these tiles
into mosaics $4\degree$ in longitude, and $16\degree$ in latitude,
centered on the Galactic plane.  Adjacent mosaics overlap in longitude
by $0\fdg5$.  Our WISE mosaics collectively cover $|b| \le 8\arcdeg$
over the entire range of Galactic longitudes.  For both photometric
bands, we use a pixel size of $2\arcsec$.  The WISE tiles are
individually background corrected so the image tile fluxes are not on
an absolute scale. We find that, when making background corrections,
the Montage software often introduces large angular-scale ($\sim
10\arcdeg$) variations in the background level.  The quality of the
mosaics is nevertheless sufficient for the identification of sites of
massive star formation.

To our knowledge, there are only five massive star formation complexes
known outside the latitude range of our WISE mosaics ($|b| \le
8\arcdeg$): Mon~R2, the California Nebula, Orion, BFS~11, and
G159.6$-$18.5 in the Perseus molecular cloud.  For these regions we
create and search WISE mosaics as before and all five are detected by
WISE.  The California Nebula (Sharpless~220) is a long filamentary
structure seen at 12\,\micron.  We include it in the catalog, although
its appearance is different from that of most sources.  The region
G159.6$-$18.5 is easily detected with WISE, although it has very weak
radio continuum emission in the GB6 survey \cite{gregory96}.
\citet{andersson00} also detected weak radio continuum emission from
this nebula.

There is no single radio continuum survey covering the entire sky that
can be used for this work.  Instead, we use a variety of radio
continuum surveys, listed in Table~\ref{tab:continuum}.  Because of
the different spatial scales probed by these surveys, it is useful to
examine all available radio data, even if these data cover the same
Galactic zone.  This is especially true in the first Galactic quadrant
where there are numerous high-quality surveys.  For example, we find
that the VGPS is the most sensitive survey extant for extended diffuse
emission and is useful for identifying large diffuse \hii\ regions,
but MAGPIS 20\,cm data \citep{helfand06} boasts a higher angular
resolution and point source sensitivity than the VGPS and is useful
for identifying ultra-compact (UC) \hii\ regions.  We encounter a
similar situation in the southern hemisphere with the SGPS and SUMSS
data sets.  The simultaneous use of both high- and low-resolution
radio data produces a more complete catalog.

We examine by eye WISE and radio continuum images spanning the entire
Galactic plane, $|b| \le 8\degree$.  As in past work
\citep{anderson11} we search for radio continuum emission spatially
coincident with objects having characteristic MIR morphologies.
  Each field was searched at least three times by one of us (LDA), in
  addition to searches by other group members.  One of us (LDA)
  determined if an object identified by one or more group members would
  be included in the WISE catalog.  This method of searching each
  field multiple times ensures a more complete catalog of
\hii\ regions and \hii\ region candidates.  For each identified source
we store in the catalog the position and approximate radius of a
circular aperture that encloses the associated MIR emission.  For
complicated regions of the Galaxy, we verify that the
  WISE sources are distinct using the {\it Spitzer} GLIMPSE
\citep{benjamin03, churchwell09} and MIPSGAL \citep{carey09} legacy
data, when possible.

Figure~\ref{fig:search} shows our procedure for a $3\arcdeg \times
1\arcdeg$ portion of the Galaxy centered at \lb = (30\arcdeg,
0\arcdeg).  The circle sizes in this figure approximate the extent of
the MIR emission associated with each source.  It is this size that is
cited in the catalog.  In the inset panels, we label regions
observed in RRL emission.  In the left inset, G031.050+00.480,
G030.956+00.599, and G030.951+00.541 were observed in RRL emission;
the other three \hii\ regions in the field are ``grouped''
\hii\ region candidates that are positionally associated with the
three known regions.  The procedure of associating \hii\ regions and
candidates with one another to create groups is described in
Section~\ref{sec:groups}.  The middle panel shows the well-known G29
\hii\ region complex.  Because of the high density of sources in this
part of the Galaxy, however, we do not associate the \hii\ region
  candidate (cyan) towards the southwest of the inset with G29.  The
  right inset shows two \hii\ regions observed in RRLs: a bright
  compact region (G028.983$-$00.603), and a more diffuse region
  (G029.094$-$00.713).  There is a compact grouped \hii\ region
candidate positionally associated with G028.983$-$00.603.  To the
north is an example of an extended radio quiet candidate, and to the
south is an example of a small radio quiet source.

In addition, we perform an automated search for \hii\ region
candidates by matching NVSS 20\,cm continuum data \citep{condon98} and
MAGPIS 20\,cm continuum data \citep{helfand06} with WISE point
sources.  To reduce the number of spurious matches, we only include
point sources that have WISE colors $[F_{12}/F_{22}] > 0.5$
\citep[][their Table~1]{anderson12a} and $F_{12} \le 15$, where
$F_{12}$ and $F_{22}$ are the 12\,\micron\ and 22\,\micron\ WISE
fluxes, respectively.  This effort only yielded another 20
\hii\ region candidates.  This suggests that the visual search alone
is sufficient to identify most \hii\ regions and \hii\ region
candidates in the Galaxy.  Unfortunately, these WISE color criteria
cannot reliably distinguish between \hii\ regions and planetary
nebulae (PNe).  The automated search therefore identified hundreds of
PNe candidates, identifiable by their extended latitude distribution
and lack of MIR nebulosity.  The NVSS does not cover the southern sky
but given the marginal results in the northern sky automated search,
we did not attempt to repeat the automated search in the south.

{\it Ex post facto} we correlate all \hii\ region candidates lacking
radio continuum emission in the surveys of Table~\ref{tab:continuum}
with the radio continuum observations of \citet{urquhart07a,
    urquhart09} at 6\,cm, \citet{sanchez-monge13} at 3.6 and 1.4\,cm,
  and the ``CORNISH'' 6\,cm VLA survey \citep{hoare12, purcell13}.
  For \citet{urquhart07a, urquhart09} and \citet{sanchez-monge13}, we
  search within radii of $30\arcsec$ and find 45, 10, and 28 matches,
  respectively.  For CORNISH, we visually examine data for sources
  identified by the CORNISH team as 
\hii\ regions\footnote{http://cornish.leeds.ac.uk/public/index.php}.
We find 12 matches with WISE sources that were previously identified
as having no radio emission or that were in confused areas of the
Galaxy.  Additionally, we add nine objects identified by CORNISH that
we missed in our visual search (all nine are detected in MAGPIS).
These nine objects are point sources in the WISE data, which
  is why they were not initially identified.
%Their
%  addition represents an increase in the candidate sample of
%  $<0.5\%$.}

PNe and galaxies also have spatially coincident MIR and radio
continuum emission so they can contaminate our sample of \hii\ region
candidates.  PNe typically have very small angular sizes.  Their MIR
colors are not distinct from those of \hii\ regions \citep[see][ and
  references therein]{anderson12a}.  In the second and third Galactic
quadrants, external star forming galaxies can look visually similar to
Galactic \hii\ regions.  In the first and forth quadrants, diffuse
background emission in the Galactic plane makes the detection of
external galaxies very difficult.  We performed a search using the
SIMBAD database\footnote{http://simbad.u-strasbg.fr/simbad/.} on all
visually and automatically identified \hii\ regions and candidates
whose emission is unresolved in the WISE data, and on all candidates
in the second and third Galactic quadrants.  These searches were
  done manually on each such object, and the results were individually
  verified.  This process removed hundreds of \hii\ region candidates
that have been identified as PNe or galaxies, especially at high and
low Galactic latitudes.

The lack of many external galaxies detected in the first and fourth
Galactic quadrants indicates that we may be less complete in the
  inner Galaxy, due to the increased shot noise from diffuse
emission.  \citet{kobulnicky13} for example find that for background
levels of 200\,MJy/sr in GLIMPSE 8.0\,\micron\ data ($\sim20\%$ of
{\rm all GLIMPSE pixel values} are higher than this), the data are only $\sim
50\%$ complete for point sources between 13th and 14th magnitude
(0.4\,mJy to 0.15\,mJy).  We can be sure that WISE has a similar
effect.  Because the WISE sensitivity limit is nearly 100 times less
  than that required to detect the faintest \hii\ region in the Galaxy
  (Figure~\ref{fig:flux_v_distance}), shot noise probably
  does not significantly impact our results.

Our catalog may not include all large diffuse \hii\ regions and young
hyper-compact \hii\ regions, although we do not believe either has
significantly affected the catalog completeness. Large \hii\ regions
have low surface brightness.  Figure~\ref{fig:flux_v_distance}, which
shows the integrated 22\,\micron\ intensity, does not account for
resolved low-surface brightness regions.  The physical size of an
\hii\ region is related to its age, ionizing flux, and nebular density
\citep[e.g.][]{spitzer}.  The largest diffuse \hii\ regions must be
both near to the Sun and also evolved.  Such regions are generally
known from H$\alpha$ surveys.  Confusion in the Galactic plane,
however, may limit our ability to detect them.  Young hyper-compact
\hii\ regions are optically thick at the radio wavelengths used here.
All the hypercompact \hii\ region candidates of \citet{sewilo04a} are
already included in our catalog.  Thus, even hypercompact
\hii\ regions meet our selection criteria.

\section{THE WISE CATALOG OF GALACTIC H\,{\small \bf II} REGIONS\label{sec:catalog}}
The WISE catalog of Galactic \hii\ regions contains         8399

entries for two types of object: known \hii\ regions and candidate
\hii\ regions.  There are         1523
 sources in the catalog that are
known to be \hii\ regions because they have measured RRL or H$\alpha$
emission (hereafter the ``known'' sample).  There are         1986

candidate \hii\ regions in the catalog that are spatially coincident
with radio continuum emission, but do not yet have any RRL or
H$\alpha$ observations (hereafter the ``candidate'' sample).  Our
group's previous RRL studies show that $\sim95\%$ of such targets are
\hii\ regions.  We find that          650
 of these
candidates are also positionally associated with known \hii\ region
complexes, so the probability of their being {\it bona fide} \hii\ regions
is even higher (hereafter the ``group'' sample).  At the sensitivity
limits of existing surveys,         4124
 catalog sources show no
radio continuum emission (hereafter the ``radio quiet'' sample).
There are an additional          115
 objects that lack high
quality radio continuum data.
%and therefore their classification is
%unknown.

%The WISE catalog of Galactic \hii\ regions lists \input total objects.
%There are four types of object: known \hii\ regions with measured RRL
%or H$\alpha$ emission (hereafter the ``known'' sample), {\bf grouped
%  \hii\ region candidates that have detected radio continuum emission
%  and are positionally associated with objects in the known sample
%  (the ``group'' sample)}, \hii\ regions with detected radio continuum
%emission that have not been observed in RRLs or H$\alpha$ (the
%``candidate'' sample), and \hii\ region candidates lacking radio
%continuum emission (the ``radio-quiet'' sample).  {\bf The four
%  catalog types of known, group, candidate, and radio-quiet candidate
%  are ordered with decreasing certainty that the object is an
%  \hii\ region.}  There are an additional \input noradio_data objects
%that lack high quality radio continuum data and therefore their
%classification is unknown.
%{\bf
%  All catalog sources have the characteristic MIR morphology of
%  12\,\micron\ emission surrounded by 22\,\micron\ emission.}

We give the WISE catalog of Galactic \hii\ regions in
Table~\ref{tab:catalog}.  The first seven columns list parameters
derived from our searches: the source name, the classification (``K''
for known, , ``G'' for group, ``C'' for candidate, ``Q'' for radio
quiet, and ``?''  for sources without radio data), the Galactic
longitude, the Galactic latitude, the approximate circular radius in
arcseconds required to encircle the WISE MIR emission, the
\hii\ region name, if known, and the \hii\ region group membership
name (see Section~\ref{sec:groups}).  The next seven columns give the
parameters of the RRL or H$\alpha$ observations: the Galactic
longitude and latitude of the observations, the LSR velocity and
its $1\sigma$ error, the FWHM line width and its $1\sigma$ error, and the
reference.  If there are multiple line components measured for a
source, the multiple values for the LSR velocity and FWHM line width
are separated with a semicolon.

%{\bf Sources within a group have consecutive entries in the table.
The group membership column of Table~\ref{tab:catalog} contains values
not only for group \hii\ region candidates, but also for known
\hii\ regions in the group, and any radio quiet sources that are
also positionally associated with the group.  All members of a given
group therefore share the same value in this column.

\begin{figure}[!ht]
\begin{centering}
\includegraphics[width=6.5in]{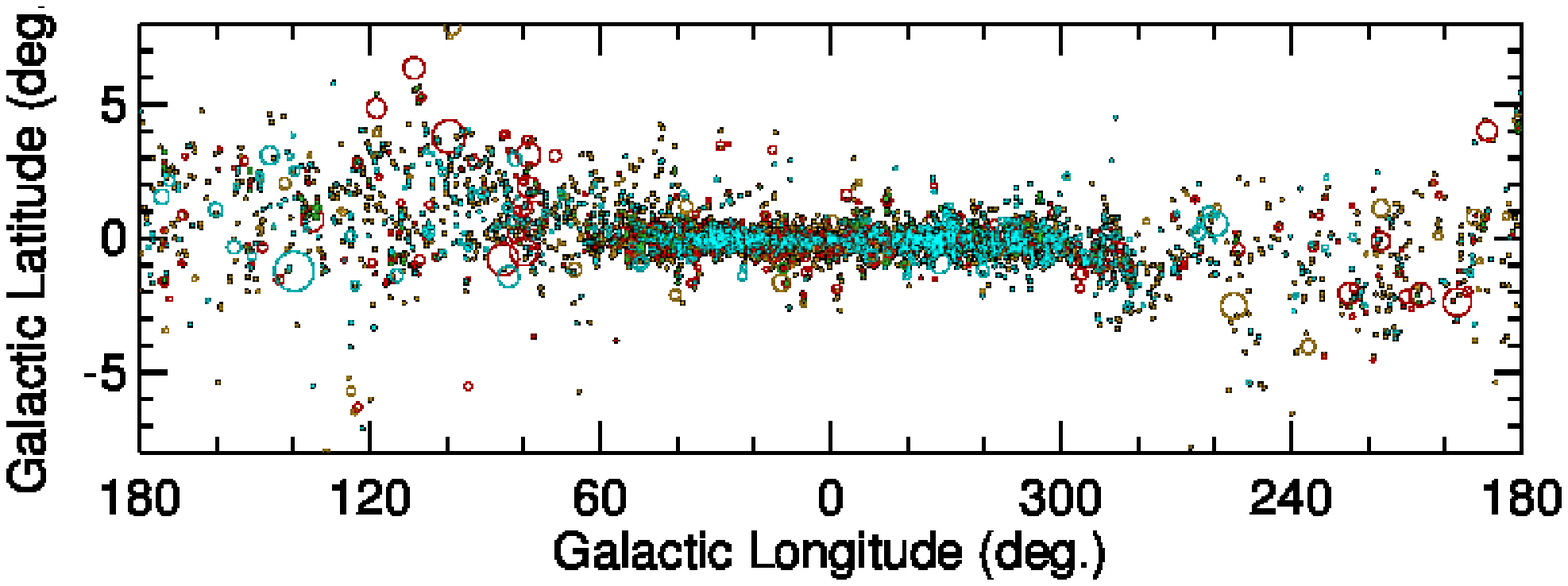}
\includegraphics[width=6.5in]{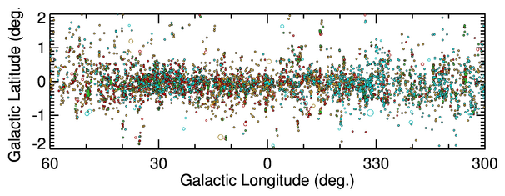}
\caption{Galactic distribution of WISE \hii\ region catalog sources.
  The red, green, cyan, and yellow circles show the locations of the
  known, group, candidate, and radio quiet sources, respectively.
The symbol sizes approximate the MIR extent of the sources.  In the
lower panel the view is restricted to $60\arcdeg \ge \ell \ge
-60\arcdeg$, to show the high density of sources in the inner Galaxy.}
\label{fig:catalog_lb}
\end{centering}
\end{figure}

We show the Galactic locations of the catalog contents in
Figure~\ref{fig:catalog_lb}, which demonstrates the high density of
sources, especially in the inner Galaxy.  It is also clear from
Figure~\ref{fig:catalog_lb} that sources are more confined to the
Galactic plane in the inner compared to the outer Galaxy, and that
sources in the outer Galaxy are generally larger in angle.
 
\subsection{Known H\,{\small \bf II} Regions}
There are         1523
 known \hii\ regions in the catalog.  All known
\hii\ regions have measured RRL or H$\alpha$ spectroscopic emission.
To identify known \hii\ regions, we correlate the WISE-identified
sources with the catalogs of \citet{downes80}, \citet{wink82},
\citet{caswell87}, \citet{lockman89}, \citet{fich90},
\citet{lockman96}, \citet{araya02}, \citet{watson03},
\citet{sewilo04b}, \citet{anderson11}, and \citet{bania12}.  All
except for the work of \citet{fich90}, which measured H$\alpha$, are
RRL surveys.  If a source has been observed by multiple authors, we
use the line parameters from the most recent author, since more
recent observations are generally more reliable due to advances in
instrumentation and telescope design.  The exception to the above rule
is that we include the H$\alpha$ observations of \citet{fich90} if
there are no other observations available.  For 198 positions
  observed in RRL or H$\alpha$ emission, the telescope beam contains
  multiple, separate \hii\ regions.  In such cases for all
  \hii\ regions within the telescope beam we list the same RRL or
  H$\alpha$ observational parameters.

This WISE-identified catalog improves upon the effort of
\citet{paladini03} in that it includes more recent observations,
removes sources duplicated by multiple authors, updates the positions
using high-resolution MIR data, and removes contaminants now known not
to be \hii\ regions.  Although they have detected RRL or
  H$\alpha$ emission, we exclude 133 targets previously identified as
  \hii\ regions (Table~\ref{tab:removed}).  About half of these are
not distinct \hii\ regions, but rather are positions observed within
large \hii\ regions.  We only use the line parameters from one
observed position per WISE source: the location closest to the nominal
source centroid.  The characteristic IR morphology is absent for 35
previously observed positions.  The detection of thermal RRL emission
in the direction of these sources is likely due to diffuse Galactic
plasma, which is prevalent in the inner Galaxy \citep[e.g.,][for
  discussion of emission near $\ell = 30\arcdeg$]{anderson11}.
Finally, we exclude objects that more recent observations have
  shown to be supernova remnants, PNe, or stars (see
  Table~\ref{tab:removed}).

%This same procedure
%  was followed by \citet{anderson09b} to create their catalog of
%  Galactic \hii\ regions.  }

\subsection{Groups\label{sec:groups}}
Star formation is a clustered phenomenon and many well-known
\hii\ regions are composed of multiple, individual \hii\ regions, e.g.,
Sgr~B2, W51, and G29.  Numerous less well-studied star forming
regions, however, also contain multiple, distinct \hii\ regions.  For
example, Sharpless~104 has a small compact \hii\ region on its
eastern border \citep[see][]{deharveng03}, but this compact region has
not been measured in RRL or H$\alpha$ emission and therefore does not
meet our criteria for inclusion in the known sample.  In general, only
the brightest object in an \hii\ region complex has been observed in RRL or H$\alpha$
emission.

\begin{figure}[!ht]
\begin{centering}
\includegraphics[width=6.52in]{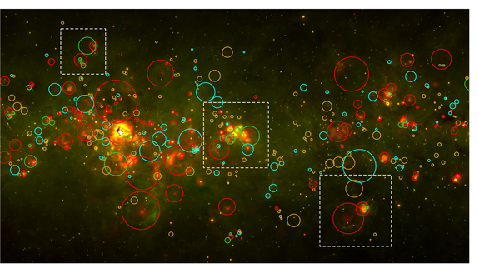}
\includegraphics[width=2.13in]{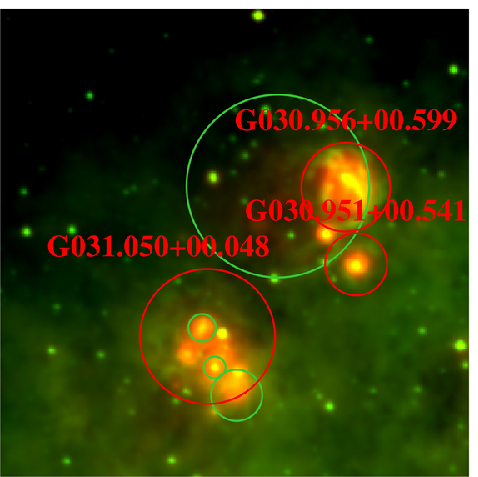}
\includegraphics[width=2.13in]{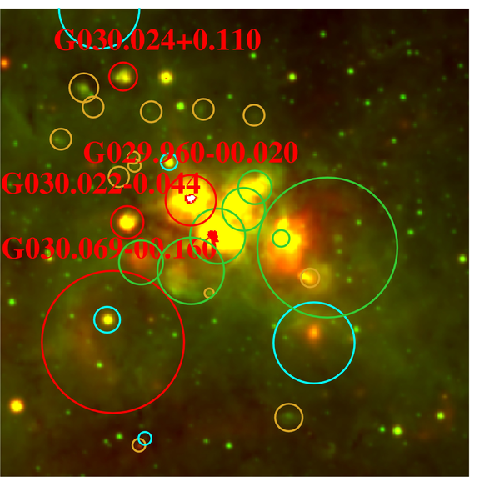}
\includegraphics[width=2.13in]{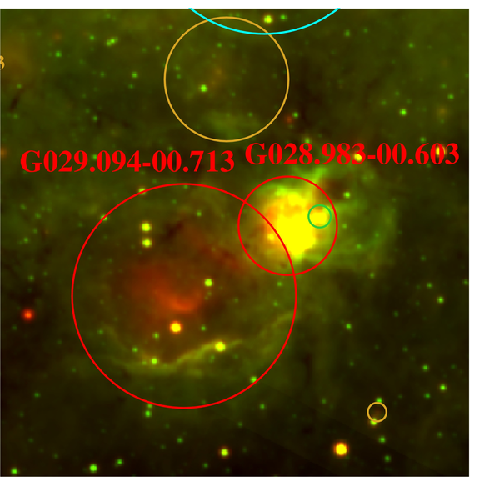}
\caption{\hii\ region search methodology.  The background images show
  WISE 22\,\micron\ data in red and WISE 12\,\micron\ data in green.
  The red, green, cyan, and yellow circles show the locations of
    the known, group, candidate, and radio quiet sources,
    respectively.  The circle sizes approximate the extent of the MIR
  emission.  The top image is $3\arcdeg \times 2\arcdeg$, centered at
  \lb = (30\arcdeg, 0\arcdeg).  The left inset is $20\arcmin$ square
  centered at \lb = (30.02\arcdeg, 0.57\arcdeg), the middle inset is
  $30\arcmin$ square centered on the G29 complex at \lb =
  (29.91\arcdeg, $-$0.06\arcdeg), and the right inset is $30\arcmin$
  square, centered at \lb = (29.03\arcdeg, $-$0.62\arcdeg).  Sources
  observed in RRL emission are labeled.}
\label{fig:search}
\end{centering}
\end{figure}

We group positionally associated \hii\ region candidates with sources
from the known sample.  In general, for group membership we require
that an \hii\ region candidate be located on, or interior to, the
photodisociation region (PDR) of a known \hii\ region (as seen at
12\,\micron), or that a known \hii\ region be located on the PDR of an
\hii\ region candidate.  We relax this criterion slightly in the outer
Galaxy and away from the Galactic plane, where there are few
\hii\ regions along a given line of sight.  For $120\degree < \ell
<240\degree$ and $|b| > 1\arcdeg$ we require only that that
\hii\ region group members are connected by diffuse
12\,\micron\ emission.  We also relax the PDR criterion for large,
bright \hii\ region complexes, e.g., G29 (Figure~\ref{fig:search}).
In such complexes we assume that all nearby, bright \hii\ regions that
are part of a single large radio continuum source are part of the same
\hii\ region group.  We largely refrain from assigning group
membership in complicated zones of the inner Galaxy where there may be
multiple \hii\ regions along a given line of sight.  We find that
         650
 sources that would otherwise be classified as
\hii\ region candidates are part of groups.  The largest group is that
associated with W49, which has 22 members.  Finally, there are          403
 radio quiet candidates that we associate with groups;
nevertheless, these sources are in the radio quiet rather than the
group catalog.  We show example groups in Figure~\ref{fig:search}.

%{\bf By our above definition, objects in the group sample have MIR
%  emission spatially coincident with radio continuum emission.  In
%  Table~\ref{tab:catalog}, we also list the group membership for
%  radio-quiet candidates, although these objects are not in the group
%  sample.}

We caution that group membership is based solely on the positional
correlation between known and candidate \hii\ regions.  This is a
necessarily subjective process.  Our hypothesis is that all group
members are physically related, but we cannot be sure without
spectroscopic observations.  We attempt to assess the reliability of
these group associations in Section~\ref{sec:molecular}, but we defer
a more detailed study to a future publication.

\subsection{H\,{\small \bf II} Region Candidates}
There are         1986
 \hii\ region candidates that have the
characteristic \hii\ region MIR morphology spatially coincident with
detected radio continuum emission but lack RRL or H$\alpha$
observations.  These are ideal targets for followup spectroscopic
  observations, which would determine if the candidates are true
  \hii\ regions.  In the GBT HRDS, \citet{anderson11} detected the
RRL and radio continuum emission from 95\% of similarly-identified
candidates.  The detection of RRL and radio continuum emission in most
cases proves that the source is a true \hii\ region, although PNe also
have RRL emission \citep{garay89, balser97}.  We have begun a program
to extend the HRDS using WISE candidates and we again find that
approximately 95\% of such candidates are detected in RRL emission.
We therefore suggest that essentially all such \hii\ region candidates
are {\it bona fide} \hii\ regions.

\subsection{Radio Quiet H\,{\small \bf II} Region Candidates\label{sec:radio-quiet}}
There are         4124
 radio quiet sources that lack detected
radio continuum emission.  This sample contains objects that have the
MIR appearance of \hii\ regions.
%and there are no objects added from
%our automated search.  
Membership in this sample is set by the sensitivity limits of existing
radio continuum surveys.  For example, data from the extremely
sensitive MAGPIS 20\,cm continuum survey allowed us to identify 162 \hii\
region candidates that would otherwise be classified as radio quiet 
based on the poorer sensitivity of all other continuum surveys.

The exact nature of the radio quiet sources is not clear.  A WISE
catalog entry may be classified as radio quiet if it contains only
intermediate mass stars, is an \hii\ region (or young stellar object)
in an early stage of its evolution, or is an \hii\ region in a late
stage of its evolution.  Intermediate mass stars may lack the
ultra-violet photons necessary to create an \hii\ region that can be
detected in the radio continuum surveys we used.  Very young
\hii\ regions are optically thick at lower radio frequencies, which
may limit our ability to detect their radio continuum emission.
Evolved \hii\ regions have low surface brightness radio continuum
emission that may similarly fall below the sensitivity limits of the
radio continuum surveys used.  The radio quiet candidate sample
therefore in all likelihood contains many different types of object.

We find that the majority of the radio quiet sources are small in
angular size and are correlated with cold dust.  This suggests that
most radio quiet sources are in the earliest phases of \hii\ region
evolution.  Over 60\% of the radio quiet sources have circular radii
$<1\arcmin$, and over 80\% have circular radii $<2\arcmin$ (these
numbers are $\sim 25\%$ and $\sim50\%$ for both the known and
\hii\ region candidates samples; see Section~\ref{sec:sizes}).  In
Section~\ref{sec:ir_continuum} we show correlations between the WISE
catalog sources and various IR, sub-millimeter, and millimeter
and catalogs, including the cold dust traced by the ATLASGAL
\citep{schuller09} and BOLOCAM Galactic Plane Survey
\citep[BGPS;][]{aguirre11} surveys.  Over two-thirds of all radio
quiet sources that have angular radii $<240\arcsec$ are associated
with ATLASGAL \citep{schuller09} or BOLOCAM Galactic Plane Survey
\citep[BGPS;][]{aguirre11} sources (Section~\ref{sec:ir_continuum}).

A smaller subset of the radio quiet sources is composed of true
\hii\ regions with weak radio continuum emission.  For example, the
bubble in Perseus, G159.6$-$18.5, is ionized by HD~278942, an
O9.5-B0~V star \citep{andersson00}.
%Rather than diffuse emission
%  that is more common for \hii\ regions, the 22\,\micron\ emission of
%  G159.6$-$18.5 is centered on the exciting star HD~278942.  This is
%  characteristic of ionization by early B-type stars (L.~Deharveng,
%  private communication).  
The distance to Perseus is only $\sim300$\,pc \citep[see][who place
  the distance closer to 200\,pc]{hirota08}.  While the Perseus bubble
is easily detected with WISE, if it were much more distant its radio
continuum emission would not be detected.  Many true
\hii\ regions more distant than Perseus would be classified as radio
quiet.

%Approximately xxx\% of these sources are larger than $1\arcmin$ in
%diameter.  Based on their similar MIR morphologies to known
%\hii\ regions, we hypothesize that essentially all these sources are
%true \hii\ regions, but that the intensity of their radio continuum
%emission is below the sensitivity of the extant radio continuum
%surveys.  Testing this hypothesis would require radio continuum
%observations that are very sensitive to diffuse low-intensity
%emission.

\subsection{Sources Lacking Radio Data}
We find          115
 sources for which high-quality radio
continuum data are not available.  Portions of the searched Galactic
zone are not covered by the radio continuum surveys in
Table~\ref{tab:continuum}, primarily within $|b| < 1\fdg5$, $\ell >
255\degree$.  The classification of these objects will require
new radio continuum observations.

\section{CORRELATION WITH PREVIOUS WORK}
To better characterize our WISE objects and to provide velocities for
those not yet observed in RRL or H$\alpha$ emission, we spatially
correlate our WISE sources with catalogs of dust continuum (at
  IR, sub-mm, and mm-wavelengths) and molecular line emission.  We
only attempt to correlate WISE sources that have angular radii
$<4\arcmin$.  Since nearly all dust continuum and molecular line
catalogs consist of unresolved objects, we find that restricting the
size of the WISE objects is necessary to avoid spurious matches
between large \hii\ regions and compact objects.  The $4\arcmin$
  limit was chosen because it resulted in an acceptably low number of
  spurious matches between large WISE sources and nearby compact
  objects, based on a visual inspection of the
  results.  The exception to the above size criterion is the
correlations with the larger \hii\ regions from \citet{anderson09b},
where we search all WISE sources for a match.  While each dust
continuum and molecular line survey has a different spatial
resolution, we found that uniformly using a search radius of
$1\arcmin$ produces the best correlations between the molecular line
and the ionized gas velocities.  The angular size restriction of
  $4\arcmin$ therefore allows for a maximum positional offset of 25\%
  for the largest sources studied here.  We caution that these
positional correlations likely result in some false-positives.  To
mediate this effect, we only use molecular line velocities within
10\,\kms\ of the RRL or H$\alpha$ velocity (if known).

The statistics of these positional correlations are summarized in
Table~\ref{tab:correlations} and the properties of WISE objects that
are spatially correlated with molecular line emission are given in
Table~\ref{tab:mol_lines}.  Table~\ref{tab:correlations} lists the
survey name, the approximate longitude range of the survey, the number of known, group, candidate, and radio quiet
sources matched, the total number of sources matched (including those
lacking radio continuum data), the percentage of WISE sources matched
over the dust continuum or molecular line survey area, the percentage
of survey sources matched with the WISE \hii\ region catalog, the
wavelength of the survey (for continuum), the molecule observed (for
spectral lines), and the survey reference.  The LSR velocities of
molecular line emission that is spatially correlated with WISE objects
are given in Table~\ref{tab:mol_lines}.  Listed are the WISE source
name from Table 2, the source name from the molecular line
observations, the Galactic longitude and latitude of these data, the
LSR velocity, the molecule observed, and the reference.

\subsection{Dust Continuum Emission\label{sec:ir_continuum}}
About half of the WISE sources that have angular radii $<4\arcmin$ are
positionally associated with a previously identified dust continuum
source from the Red MSX Source (RMS), ATLASGAL, BGPS, or Milky Way
Project (MWP) surveys.  For the RMS survey, we only search near
sources identified as young stellar objects (YSOs) or \hii\ regions.
These identifications were based largely on the source MIR
morphologies and on the detection of molecular gas.  That only half of
all angularly small WISE sources have been identified previously
suggests that the WISE catalog is sampling a different population of
objects compared to previous work.  

About half of all RMS YSOs have a positional match with a WISE
catalog source, whereas about three quarters of all RMS \hii\ regions have a
WISE \hii\ region catalog source.  Because the angular size
  of RMS \hii\ regions and WISE catalog \hii\ regions is potentially
  quite large, we expect a number of true correlations to be excluded
  by the 1\arcmin search criterion.  A visual examination of WISE
data indicates that they are positionally correlated nearly 100\% of
the time.

About half of all WISE sources are positionally correlated with
  an object from the ATLASGAL and BGPS surveys, over the respective
  areas of these surveys.  These are the highest percentages of all
  the dust continuum surveys.  The percentage of ATLASGAL and BGPS
  sources matched with a WISE catalog object, however, is only
  $\sim20\%$, for both surveys.  This shows that these catalogs mostly
  contain objects not positionally correlated with WISE \hii\ regions
  and candidates.

The MWP \citep{simpson12} harnessed the power of thousands of on-line
volunteers to identify MIR ``bubbles'' in Spitzer images. Here, we
search for the more general MIR characteristics of \hii\ regions,
which often cannot be characterized as a bubble \citep{anderson11}.
The MWP catalogs the properties of ellipsoids that approximate the
inner and outer radii of each bubble.  There are in fact two MWP
bubble catalogs that contain ``large'' and ``small'' bubbles,
respectively.  For a positive WISE-MWP correlation, we requre that:
(a) the difference between the WISE and MWP positions be less than
half of the MWP outer semi-major axis; and (b) the WISE angular size
be within a factor of two of the MWP outer semi-major axis.  As
expected, the correlation is quite high: approximately half of all MWP
bubbles are correlated in position with a WISE object.

\subsection{Molecular Line Observations\label{sec:molecular}}
We correlate the positions of the WISE \hii\ region catalog objects
with large molecular line surveys to provide velocities and distances
for sources lacking such information.
%We use \cor\ data from
%\citet{urquhart07b, urquhart08b} (RMS), \citet{anderson09b} (data
%  from the Galactic Ring Survey \citep[GRS;][]{jackson06}), and
%\citet{wienen12} (ATLASGAL); \co\ data from \citet{wouterloot89}; CS
%data from \citet{bronfman96} (IRAS sources); HCO$^+$ data from
%\citet{schlingman11} (BGPS); NH$_3$ data from \citet{urquhart11}
%(RMS), \citet{dunham11} (BGPS), \citet{wienen12} (ATLASGAL), and
%\citet{purcell12} (HOPS); and multiple molecular species from
%\citet{foster11} (MALT90).  
For sources that are detected in multiple
transitions, we use the molecular information from the dense gas
tracers (CS, NH$_3$, and HCO$^+$) instead of that from CO.  Ideally,
the velocities of all tracers would agree but CO often has multiple
components along the line of sight, which can make velocity
assignments challenging.  For the dense gas tracers that were observed
by multiple authors, we use the data from the most recent
observations. If the WISE source is a known \hii\ region with a single
ionized gas velocity, we only use the molecular gas velocity if it
is within 10\,\kms\ of the ionized gas velocity.

We find molecular velocities for         2262
 sources; this is
30\% of all objects with radii $<4\arcmin$.  Of these,         1667
 of these do not have RRL or H$\alpha$ velocities.  The
mean absolute difference between the molecular and ionized gas
velocities is $3.1\pm2.3$\,\kms, and the mean difference is
$-0.4\pm3.8$\,\kms.  Within the errors, the velocity differences are
identical for all the molecular species studied here.  These values
are similar to those found by \citet{anderson09b} in their study of
\cor\ associated with \hii\ regions: they found a mean absolute
difference of $3.0\pm2.4$\,\kms\ and a mean difference of
$0.2\pm3.8$\,\kms.

\begin{figure}[!ht]
\begin{centering}
\includegraphics[width=6.5in]{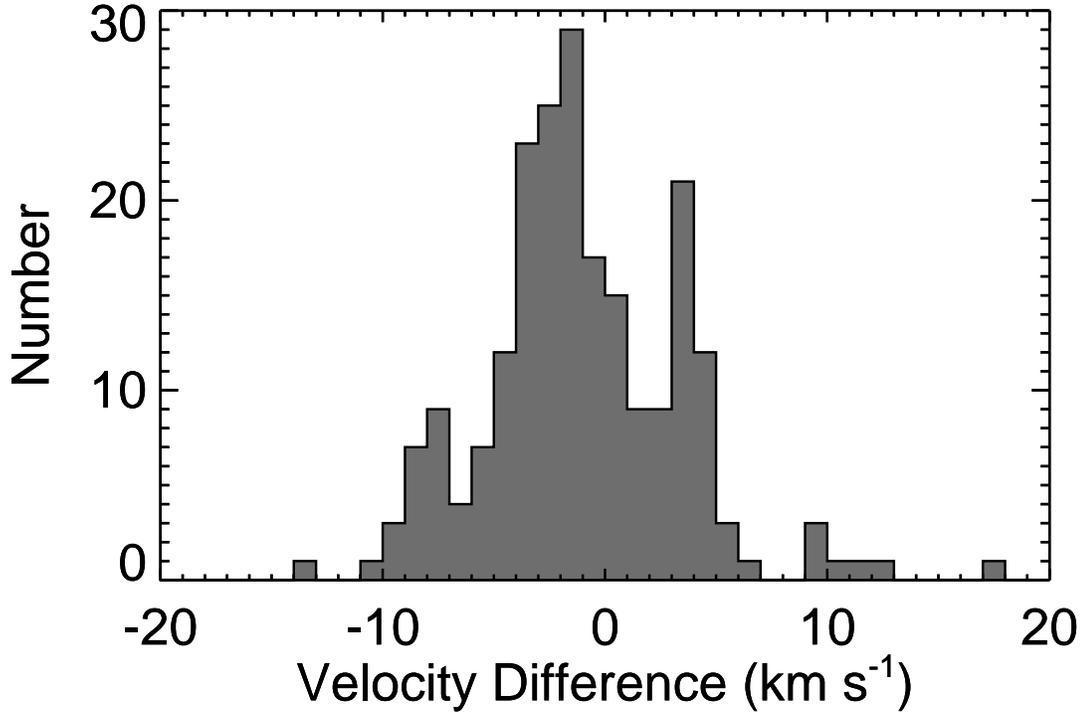}
\caption{Difference between the molecular and ionized gas LSR
    velocities for the group sample: $\Delta V = V_{\rm mol} - V_{\rm
      \hii}$. For the vast majority of WISE objects, the group sample
    associates sources with similar velocities.}
\label{fig:assoc_vlsr}
\end{centering}
\end{figure}

We defined source group associations based solely on positional
juxtapositions.  When we now add molecular velocity information, we
find that the vast majority of group members share similar LSR
velocities. This gives further support to the hypothesis that group
members are all {\it bona fide} \hii\ regions in the same star forming
complex.  We show the LSR velocity difference between the molecular
and group ionized gas emission, $\Delta V = V_{\rm mol} - V_{\rm
  HII}$, in Figure~\ref{fig:assoc_vlsr}.  The group ionized gas
velocities are those of the group member that was measured in RRL or
H$\alpha$ emission.  The means of the absolute velocity difference and
the velocity difference are $4.2\pm4.0\,\kms$ and $-1.3\pm5.7\,\kms$,
respectively. Although these velocity differences are $\sim50\%$
higher than what we find for the entire WISE sample, this is not
significant within the errors.  These statistics, as well as
Figure~\ref{fig:assoc_vlsr}, do not include 8 sources with absolute
velocity differences greater than 20\,\kms.  These objects comprise
only 4\% of the total number of grouped objects that have molecular
line velocities.  We keep these 8 sources in the group sample.

%champ barnes et al 2011
%nanten master catalog

%The sources both for the dust
%continuum and the spectroscopic surveys are all compact objects that
%are essentially point sources in their identification surveys (MSX for
%the RMS survey; IRAS for \citet{bronfman96}; 870\,\micron\ clumps for
%\citet{weinen12}, 1.1\,mm clumps for \citet{schlingman11}.  The
%candidate \hii\ regions, on the other hand, have a wide range of sizes
%from xxx to xxx, averaging xxx.  The candidates that do have such
%observations are likely young, and still in their natal molecular
%cocoons; their average sizes are xxx.  The candidates lacking such
%observations are likely more evolved; their average sizes are xxx and
%they are a distinct sample of \hii\ region candidates from those
%identified previously through other means.

%RMS entire catalog
%RMS VLA Radio
%RMS ATCA Radio
%RMS NH3
%Bolocam
%Bronfman CS
%ATLASGAL
%ATLASGAL - NH3 only

\subsection{WISE Catalog Source Distances\label{sec:distances}}
We provide distances for         1459
 WISE catalog sources.  There
are three main ways that distances have been determined for
\hii\ regions: through their association with a maser that has a
measured parallax, spectrophotometrically, and kinematically.  The
most accurate distances for star forming regions are from
trigonometric parallaxes of masers, but there are relatively few such
parallaxes known.  To make such a measurement, a star forming region
must have a bright associated maser, and many massive star formations
do not meet this criterion.  Spectrophotometric distances require one
to accurately identify the main ionizing source of an \hii\ region,
and to determine its spectral type.  Because spectral types can be
more readily determined for bright stars that have a low line-of-sight
extinction, in general spectrophotometric distances are only known for
nearby \hii\ regions.  Kinematic distances are in principle available
for all \hii\ regions with measured velocities, across the entire
Galactic disk.  \citet{anderson12c} found that kinematic distance
errors are $\sim10-20\%$ for most of the Galactic disk.  They
estimated these uncertainties from the combined effects of different
rotation curves, streaming motions, and uncertainties in the Solar
circular rotation parameters \citep[see also][]{gomez06}.  Kinematic
distances do, however, have large systematic uncertainties in parts of
the Galaxy, especially where Galactic rotation carries objects tangent
to our line of sight.

Here, we only use maser parallax and kinematic distances.  While
spectrophotometric distances can be more accurate than kinematic
distances, in practice they often have large uncertainties
\citep[see][]{russeil07}.  We use parallax distances when possible (62
WISE sources), and kinematic distances otherwise.  We compute all
distances using the \citet{brand86} rotation curve model, with the Sun
8.5\,kpc from the Galactic center and a Solar circular rotation speed
of 220\,\kms.

Sources in the inner Galaxy suffer from the kinematic distance
ambiguity (KDA), which complicates the computation of kinematic
distances.  In the Inner Galaxy, for each line of sight there are two
possible distances, known as ``near'' and ``far,'' that produce
identical LSR velocities.  The tangent point distance is halfway
between the near and far distances.  Sources at the tangent point
distance do not have a KDA.  Sources beyond the Solar orbit in the
outer Galaxy also do not have a KDA.

A kinematic distance ambiguity resolution, KDAR, requires auxiliary
data to determine if the near or far distance is correct for a given
source.  Usually the auxiliary data are \hi\ or H$_2$CO.  There are
three main methods for resolving the KDA: 
\hi\ Emission Absorption (\hi\,E/A), H$_2$CO absorption, and
\hi\ self-absorption (\hi\,SA).  In the \hi\,E/A method, foreground
\hi\ absorbs the broadband \hii\ region continuum radiation.  Neutral
\hi\ gas between the \hii\ region and the observer absorbs the thermal
continuum if the brightness temperature of the \hi\ is less than that
of the \hii\ region at 21\,cm.  Because the continuum emission is not
limited to a particular frequency (velocity), all foreground \hi\ has
the potential to absorb the \hii\ region continuum.  \hii\ regions at
the near distance will show \hi\ absorption up to the source velocity,
while those at the far distance will show \hi\ absorption up to the
tangent point velocity.  In the H$_2$CO absorption method, cold
foreground molecular clouds containing H$_2$CO may absorb the
broadband \hii\ region continuum radiation.  This method is analogous
to \hi\,E/A, with H$_2$CO replacing \hi\ as the absorbing species.  In
the \hi\,SA method, cold foreground \hi\ at the near kinematic
distance absorbs line emission from warmer background \hi\ at the
far kinematic distance.  The cold foreground \hi\ is often associated
with molecular gas.  The utility of this method for \hii\ region
studies relies on the association between the \hii\ region and
molecular gas \citep[see][]{anderson09a}.  If the cold \hi\ (together
with the molecular gas and the associated \hii\ region) is at the near
distance, there will be \hi\ absorption at the velocity of its
emission; this absorption will be absent if it is at the far distance.

These methods have their strengths and weaknesses.  The \hi\,E/A
method is generally more reliable than the other methods, because
there is an \hi\ cloud on average every 0.7\,kpc along the line of
sight \citep{bania84}, compared to 2.9\,kpc for ${\rm H_2CO}$
\citep{watson03}.  Near distances found using H$_2$CO absorption are
less reliable than those of the \hi\,E/A method because a lack of
absorption may simply be caused by a lack of H$_2$CO clouds.  If ${\rm
  H_2CO}$ absorption is detected between the source velocity and the
tangent point velocity, however, these KDARs are more reliable than
those of \hi\,E/A because background fluctiations in \hi\ data can
more easily create false absorption signals.  The \hi\,SA method is
considerably less reliable than the \hi\,E/A method for \hii\ regions
\citep{anderson09a}.

Based on these considerations, we first use far distance H$_2$CO KDARs
when possible, then the \hi\,E/A KDARs, and, finally, the \hi\,SA
KDARs if no other KDAR is known. Interferometric \hi\,E/A experiments
are more reliable than those using single-dish radio data due to
decreased sensitivity to background emission fluctuations. We
therefore use the interferometric KDARs of \citet{urquhart11} whenever
possible.  In all other cases we use the most recent KDAR
determination for a source.  \citet{anderson09a} discuss sources
  with conflicting KDA determinations, which comprise a small percentage of
  the total population.  All sources having LSR velocities within
10\,\kms\ of the tangent point velocity are given the tangent point
distance.  All inner Galaxy Sharpless \hii\ regions are given the near
distance if no other distance is available.  Finally, we assign all
sources from $74\degree < \ell < 86\degree$ that have LSR velocities
between $-$15\,\kms\ and 15\,\kms\ to the Cygnus complex, using the
maser parallax distance of DR\,21 \citep{rygl12}.

Under the assumption that the WISE catalog objects are associated with
the objects targeted in the molecular line surveys of
Section~\ref{sec:molecular}, we provide distances to          641
 WISE \hii\ region
candidates and radio quiet sources.  We use \hi\,SA and infrared dark
cloud association KDARs from \citet{dunham11} (of BGPS objects) and
the \hi\,SA KDARs from \citet{urquhart12} (of RMS objects).  These
studies use the same positions from the BGPS and RMS surveys,
respectively, so no additional positional matching with WISE sources
is required.  As with the molecular gas associations, we assume that
the KDARs for objects within $1\arcmin$ of the WISE sources also apply
to the WISE sources themselves.  

%There are xxx \hii\ regions with spectrophotometric distances in the
%\citet{russeil98} compilation and \citet{moises11} catalogs.  For
%\hii\ regions without known maser parallaxes, we use these
%spectrophotometric distances.  We prefer the newer distances from the
%\citet{moises11} catalog to that of \citet{russeil98} and use these if
%distances exist for a given region from both authors.
%\citet{moises11} use two extinction laws when computing their
%distances: \citet{mathis90} and \citet{stead09}.  These represent two
%extremes and we take the published average distance found using these
%two extinction laws.

We give the distances to our WISE catalog sources in
Table~\ref{tab:distances}, which lists the WISE source name from
Tables~\ref{tab:catalog} and \ref{tab:mol_lines}, the near kinematic
distance, the far kinematic distance, the tangent point distance, the
Galactocentric radius ($R_{\rm Gal}$), the tangent point velocity
(V$_{\rm T}$), the KDAR (``N'' for near, ``F'' for far, ``T'' for
tangent point, and ``OG'' for outer Galaxy), the heliocentric distance
($d_\sun$), the uncertainty in heliocentric distance ($\sigma_d\sun$),
the Galactic azimuth in clockwise degrees from the line connecting the
Galactic center to the Sun, the height above the plane ($z =
b\,sin(d_\sun)$, where $b$ is the Galactic latitude), the method used
to derive the distance, and the reference.  For maser parallax
distances, we recompute the Galactocentric radius based on the derived
distance.

\subsubsection{Distance Uncertainties\label{sec:distance_uncertainties}}
Here, the uncertainty estimates are based upon the \citet{anderson12c}
analysis, expanded to include the entire Galaxy.  Each uncertainty
estimate is therefore computed from the combined uncertainties from
the choice of Galactic rotation curve, streaming motions of 7\,\kms,
and a change to the Solar circular rotation speed, $\Theta_0$, for a
range of LSR velocities and Galactic longitudes.  While
\citet{anderson12c} computed the effect of a change to $\Theta_0 =
250\,\kms$ \citep{reid09b}, we here use 240\,\kms\ \citep{brunthaler11}.  The
analysis is otherwise unchanged except for how we estimate the
distance uncertainty for tangent point sources.  For these we estimate
the uncertainty by adding in quadrature the source near distance
uncertainties and the difference between the near and the far
distances: $\sigma_{\rm DTP} = [(\sigma_{\rm DN})^{2} + (D_{\rm N} -
  D_{\rm F})^{2}]^{0.5}$.  This formulation takes into account the
uncertainties in any kinematic distance due to both the source
\lv\ location, the $\sigma_{\rm DN}$ term, and also the choice of
the near, far, or tangent point distances, the $(D_{\rm N} - D_{\rm
  F})$ term.

We give an estimate for the uncertainty in each distance,
$\sigma_{d\sun}$, in Table~\ref{tab:distances}.  We do not list
distances for sources that have fractional uncertainties,
$\sigma_{d\sun} / d_\sun$, exceeding 50\%.  Additionally, we do not
provide distances for sources within $10\arcdeg$ of the Galactic center
and $20\arcdeg$ of the anti-center because kinematic distances are
very uncertain in these regions of the Galaxy.  In the second and third
Galactic quadrants, the distance uncertainties average $\sim30\%$,
whereas they average $\sim15\%$ in the first and fourth quadrants.

\section{DISCUSSION\label{sec:discussion}}
\subsection{Galaxy-wide Star Formation\label{sec:starform}}
The WISE Catalog of Galactic \hii\ Regions is the most complete catalog
yet of regions forming massive stars in the Milky Way.  It provides a
large sample of objects that will enable future studies of the
structure, dynamics, and chemical evolution of the Galaxy.  Here, we
use the catalog to assess the Galaxy-wide distribution on the sky of
nebulae forming massive stars. In Table~\ref{tab:quadrants} we show the distribution of
the catalog sources for various Galactic zones. For each sample
population in the catalog, we list the percentage of that population
that resides in each Galactic quadrant, as well as in a range of
angular distances from the Galactic center and the Galactic plane.  We
also give the same information for the entire catalog and for the
molecular gas distribution. We characterize the molecular gas
distribution using the \citet{dame01} survey of \co\ emission.
We normalize the CO percentages by the integrated line intensity
summed over the sky zone of the the WISE \hii\ region catalog (see
Section~\ref{sec:sample}).

\subsubsection{The Galactic Longitude Distribution}
\begin{figure}[!ht]
\begin{centering}
\includegraphics[width=6.5in]{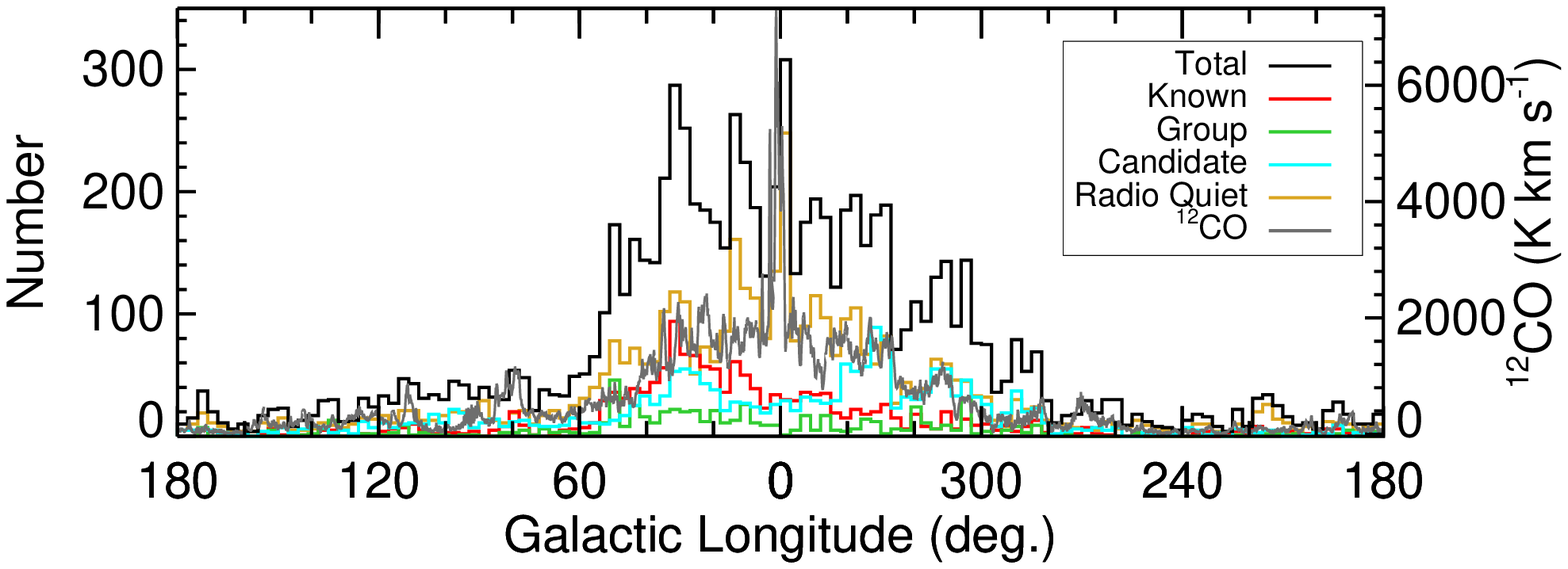}
\includegraphics[width=6.5in]{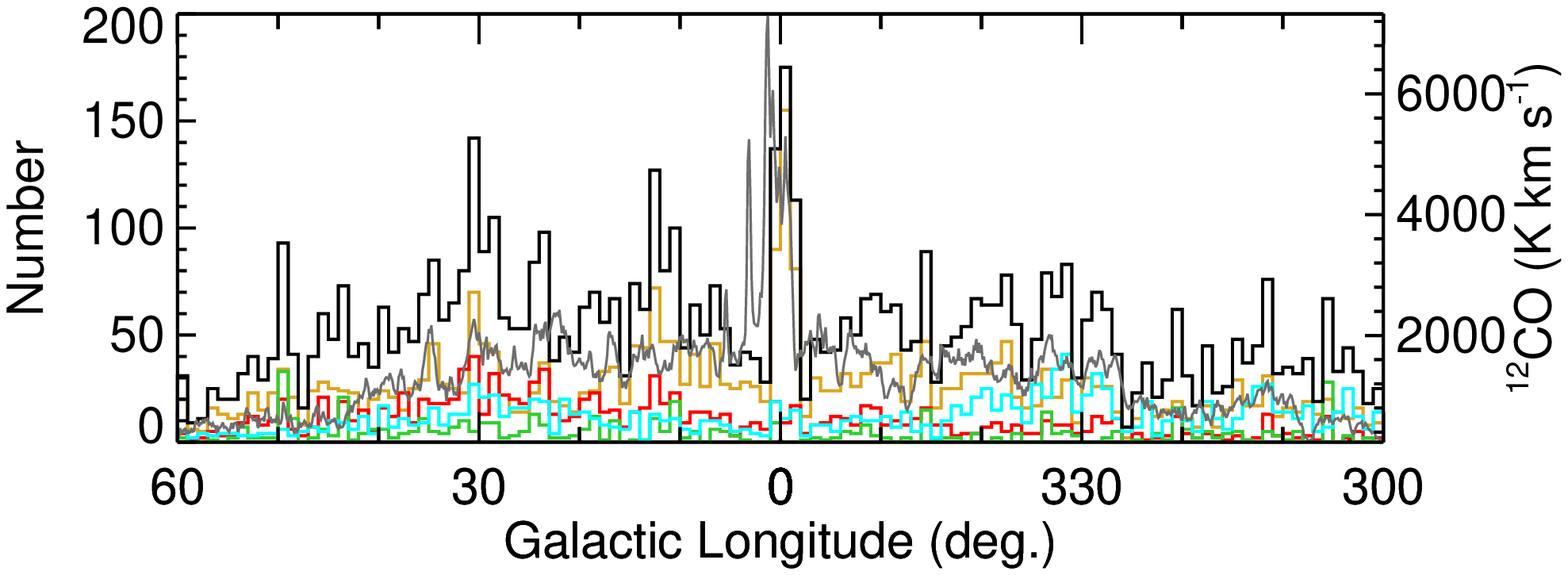}
\caption{Galactic longitude distribution of WISE \hii\ region sources.
  The red, green, cyan, and yellow histograms show the distribution for
  known, group, candidate, and radio quiet sources, respectively, whereas
  the black curve shows their sum.  The grey line is the integrated intensity of \co\ emission
  \citet{dame01}, integrated over the latitudes covered in the WISE
  \hii\ region catalog.  The Galactic center is deficient in star
  formation relative to the amount of available molecular gas.  It is
  also apparent that the fourth Galactic quadrant has many more
  \hii\ region candidates compared to the first quadrant, and that
  most of the known \hii\ regions are in the first quadrant.}
\label{fig:catalog_hist}
\end{centering}
\end{figure}

\begin{figure}[!ht]
\begin{centering}
\includegraphics[width=4in]{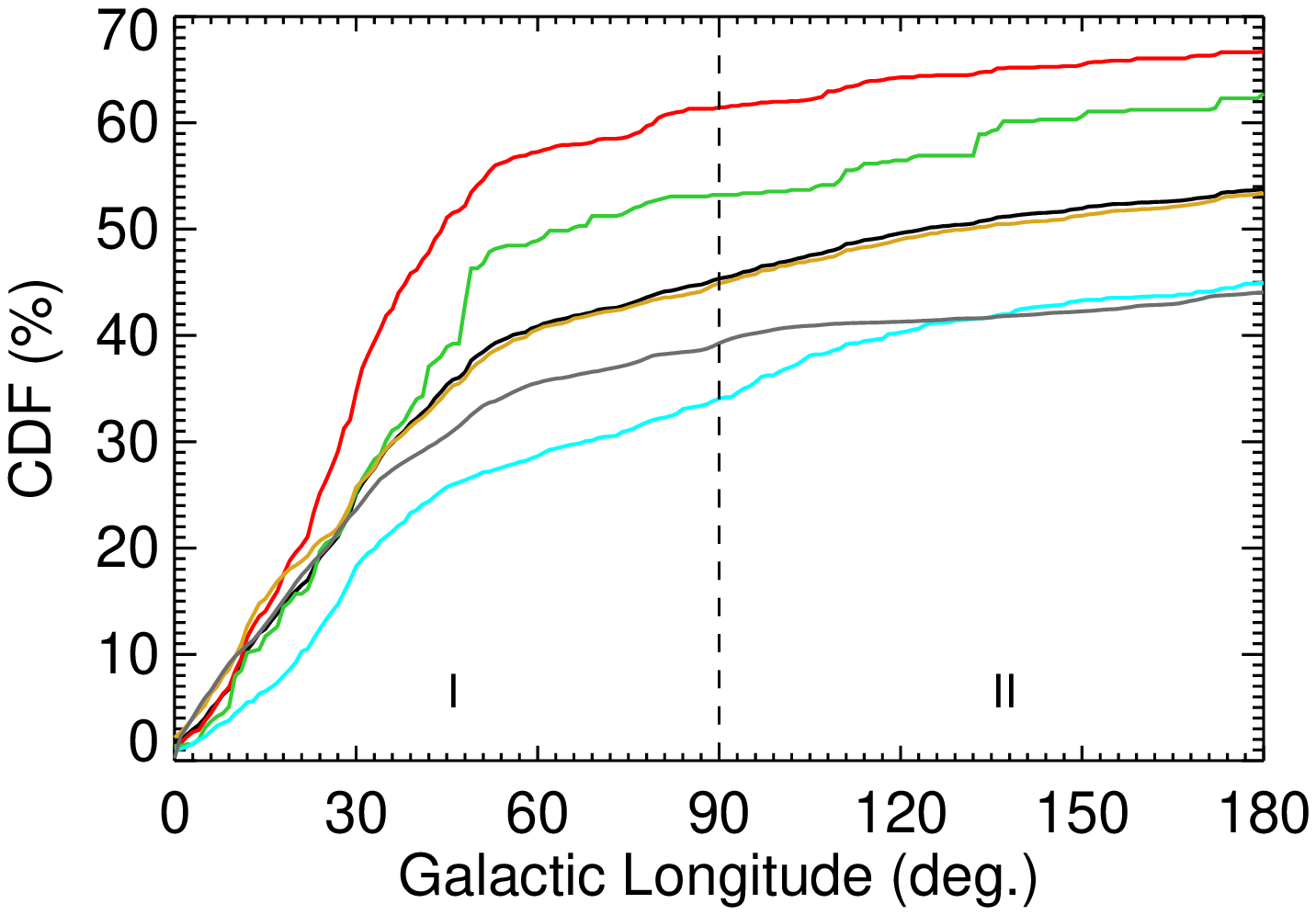}
\includegraphics[width=4in]{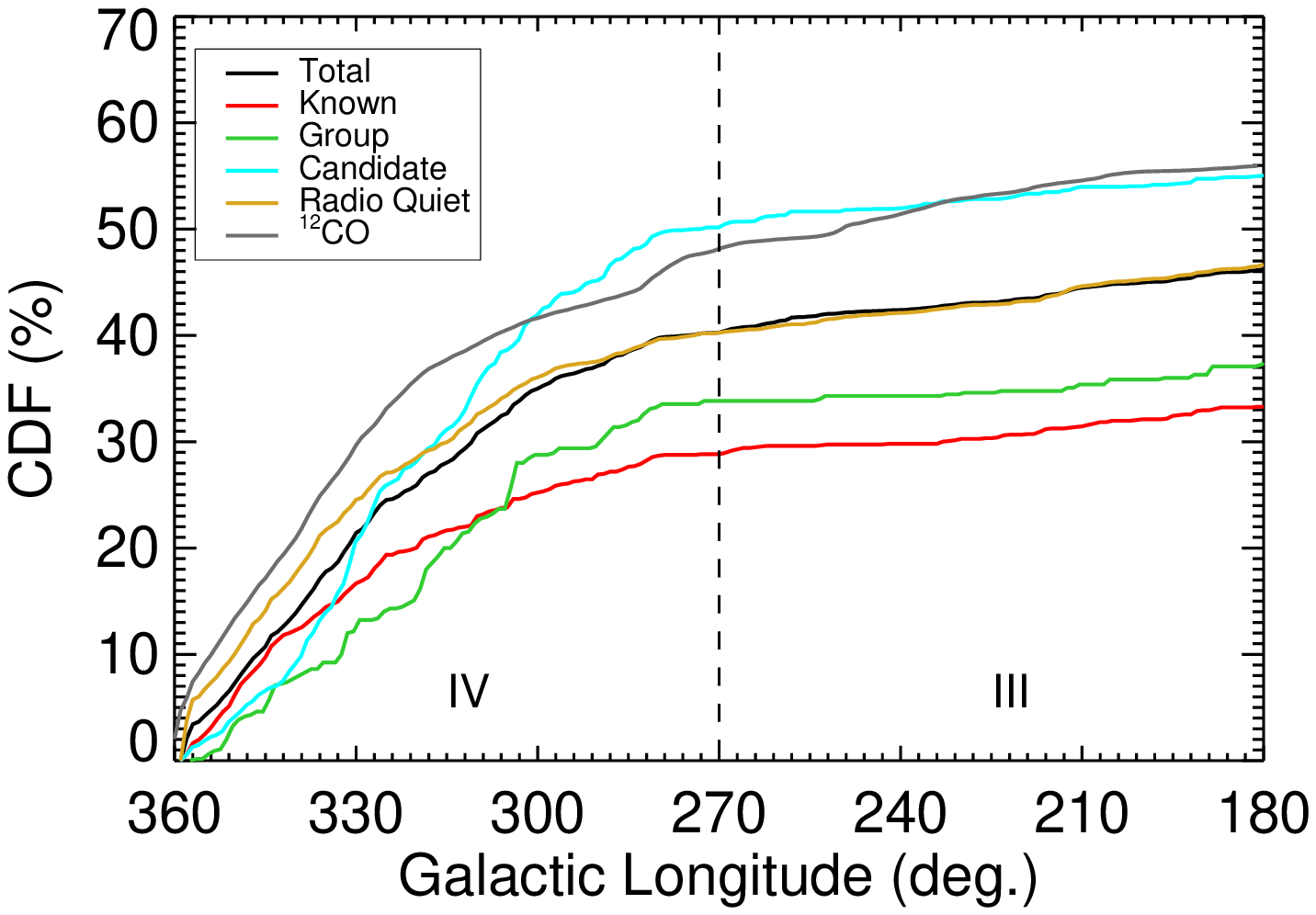}
\caption{Galactic longitude cumulative distribution functions
    (CDFs) for WISE \hii\ region catalog sources.  Shown are the CDFs
    for Galactic quadrants I and II (top) and III and IV (bottom).
    The total number of sources within each type (e.g., ``known'') for all
    longitudes is used when deriving the CDF.  The majority of sources
    reside in Galactic quadrants I and IV toward the inner Galaxy.
    The asymmetry between the first and fourth quadrants is evident,
    especially for the known and candidate samples.}
\label{fig:glong_cumulative}
\end{centering}
\end{figure}

We show the distribution in Galactic longitude for each sample
population in Figure~\ref{fig:catalog_hist}, together with the
\co\ integrated intensity from \citet{dame01} summed over latitudes
within $8\arcdeg$ of the Galactic plane.  The density of sources
increases dramatically in the inner Galaxy, as expected.  This is also
seen in the Galactic longitude cumulative distribution functions
(CDFs) that we plot in Figure~\ref{fig:glong_cumulative}.
Approximately 85\% of all sources are in the first and fourth Galactic
quadrants, and 76\% are within $60\arcdeg$ of the Galactic center.
These percentages are essentially the same for \co\ (88\% and 77\%,
respectively).  Only 15\% of catalog sources are located in the second
and third Galactic quadrants (9\% and 6\%, respectively).  Again, we
find similar percentages for \co\ (8\% and 5\%, respectively). Only
about 9\% of all known \hii\ regions are in these two quadrants.

There is, however, a north-south asymmetry in the WISE catalog of
Galactic \hii\ regions, most apparent in
Figure~\ref{fig:catalog_hist}.  About 45\% of all sources are located
in the first Galactic quadrant, and 40\% are in the fourth.  The cause
of this asymmetry may be due in part to the higher-quality radio
continuum data in the north.  When performing the visual search,
compact objects with no nebulosity would be excluded from the catalog
if they also lack radio continuum emission, because they are
indistinguishable from stars.  Because of the lower quality radio
continuum data in the south, the northern hemisphere may have more
compact catalog objects.  
%In support of this, the average size of WISE
%\hii\ region catalog sources in the north is $75\arcsec$, while it is
%$80\arcsec$ in the south.  
The opposite trend is seen for both the \hii\ regions identified in
the RMS survey (41\% in the north and 50\% in the south), and also the
\citet{bronfman96} IRAS sources that have colors indicative of high
mass star formation (36\% in the north and 43\% in the south).  Why
the \hii\ region distribution is different is unknown.  It does,
however, agree with the \co\ distribution.

There are even larger asymmetries in the known and candidate samples.
Table~\ref{tab:quadrants} shows that 61\% of all known \hii\ regions
are in the first quadrant, versus 29\% in the fourth.  About 50\% of
all \hii\ region candidates are in the fourth Galactic quadrant,
versus 34\% in the first.  This is also apparent in
Figure~\ref{fig:glong_cumulative}, as the rise in the known sample
distribution is steep in the first quadrant, and that of the candidate
sample is steep in the fourth quadrant.  Compared to the first
quadrant, the fourth quadrant has received little attention from RRL
studies.
%A fourth quadrant HRDS-like survey could reverse these
%findings.

\subsubsection{The Galactic Center}
Star formation is deficient in the Galactic center relative to the
amount of molecular material available \citep[][see also \citet{beuther12, simpson12}]{longmore13}.
While the percentages of the total integrated \co\ emission and WISE
sources detected within $60\arcdeg$ of the Galactic center are
similar, toward the Galactic center there is more \co\ emission
relative to the number of WISE sources.  Within $2\arcdeg$ of the
Galactic center, we would expect roughly twice the number of WISE
sources found, based on the \co\ intensity (see
Figure~\ref{fig:catalog_hist}).  Furthermore, the WISE sources that
lie in the Galactic center are mostly radio quiet; they may not be
true \hii\ regions.

\subsubsection{The Galactic Latitude Distribution}
\begin{figure}[!ht]
\begin{centering}
\includegraphics[width=6.5in]{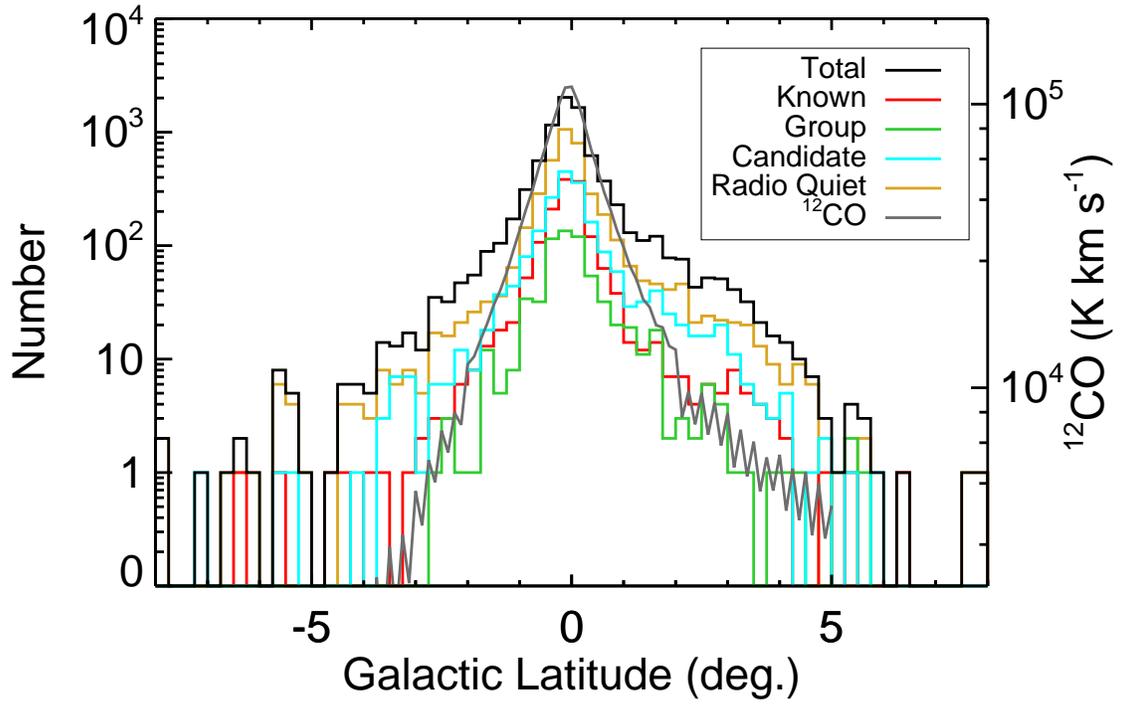}
\caption{Galactic latitude distribution of WISE \hii\ region catalog
    sources.  All samples, in addition to \co, share a similar
  distribution.  There are more \hii\ regions and candidates at
  positive Galactic latitudes than at negative Galactic latitudes.}
\label{fig:catalog_glat}
\end{centering}
\end{figure}

\begin{figure}[!ht]
\begin{centering}
\includegraphics[width=4in]{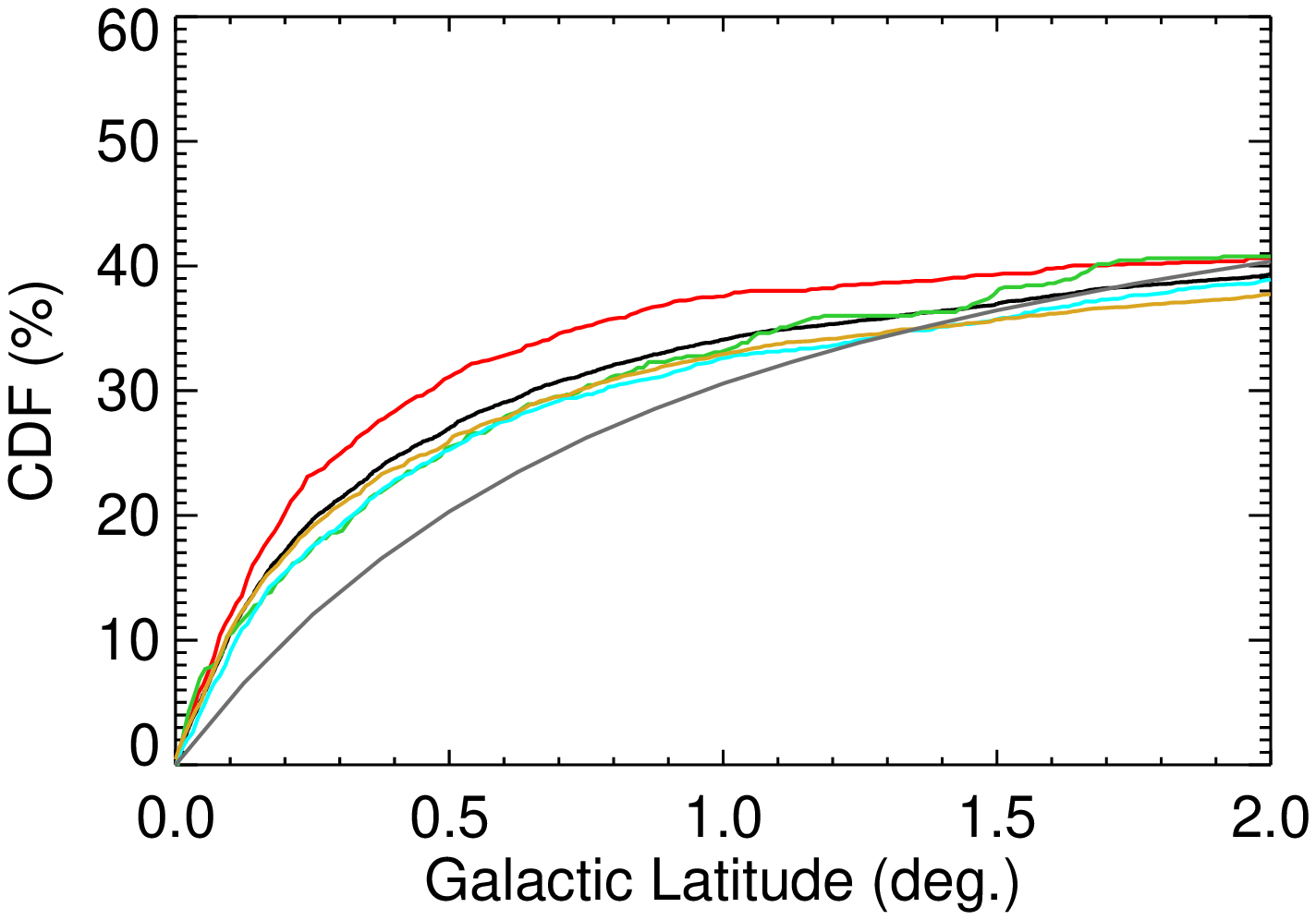}
\includegraphics[width=4in]{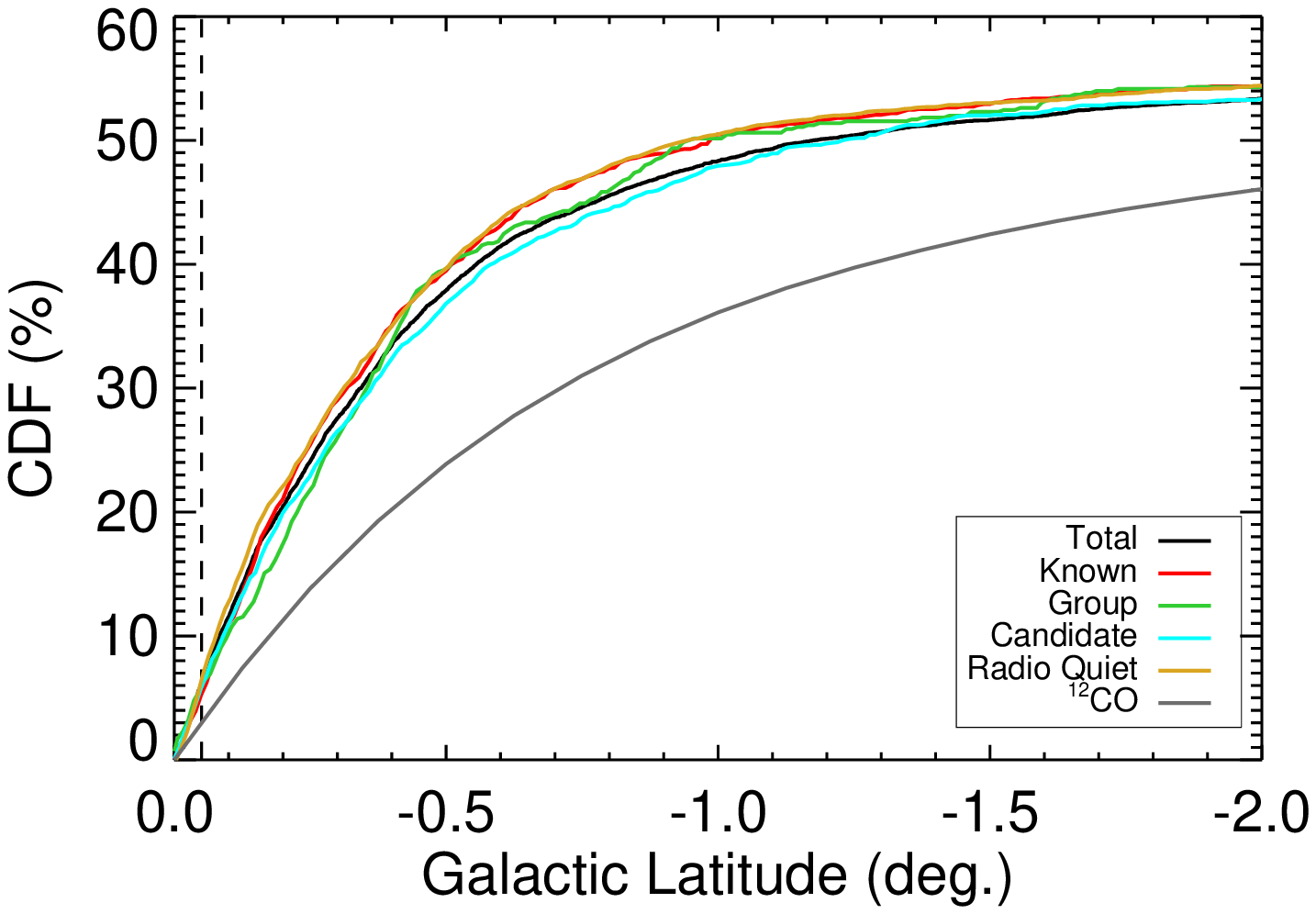}
\caption{Galactic latitude cumulative distribution functions for
    WISE \hii\ region catalog sources.  Shown are the CDFs for
    positive (top) and negative (bottom) Galactic latitudes starting
    from the nominal Galacic plane.  The total number of sources
    within each type (e.g., ``known'') for all latitudes is used when
    deriving the CDF.  The median of the entire ditribution is at $b =
    -0.05,$ marked with a vertical dashed line.  The excess of known
    \hii\ regions between $\sim 0.1\arcdeg$ and $1\arcdeg$ is
    apparent, as is the broader distribution for CO relative to that
    of the WISE catalog sources.}
\label{fig:glat_cumulative}
\end{centering}
\end{figure}

We show the distribution in Galactic latitude for each sample
population in the catalog in Figure~\ref{fig:catalog_glat}, together
with the \co\ integrated intensity. As expected, the source
distribution is strongly peaked at the nominal Galactic plane,
$b=0\arcdeg$.  This is also seen in the Galactic latitude cumulative
distribution functions (CDFs) that we plot in
Figure~\ref{fig:glat_cumulative}.  All samples have essentially the
same distribution although the \co\ distribution is considerably
broader (Table~\ref{tab:quadrants}).  There is a significant number of
sources at high and low Galactic latitudes, with $\sim20\%$ lying
outside of $|b| \le 1\arcdeg$.

All samples have an asymmetry such that there are more sources at
negative latitudes.  Over 56\% of all sources are at negative
latitudes and 44\% are at positive latitudes. The median of the
distribution is at $b = -0.05\arcdeg$, a value that holds for all
samples and also for \co\ (see Figure~\ref{fig:glat_cumulative}).  The
cause of this offset is generally ascribed to the Sun lying above the
true Galactic mid-plane.  There is an especially high density of known
\hii\ regions between $\sim 0.2\arcdeg$ and $1\arcdeg$, relative to
the other samples.  This is likely a result of the fact that most
known \hii\ regions are in the first Galactic quadrant, which on the
whole has a positive latitude offset.

\subsection{Angular Sizes \label{sec:sizes}}
\begin{figure}[!ht]
\begin{centering}
\includegraphics[width=6.5in]{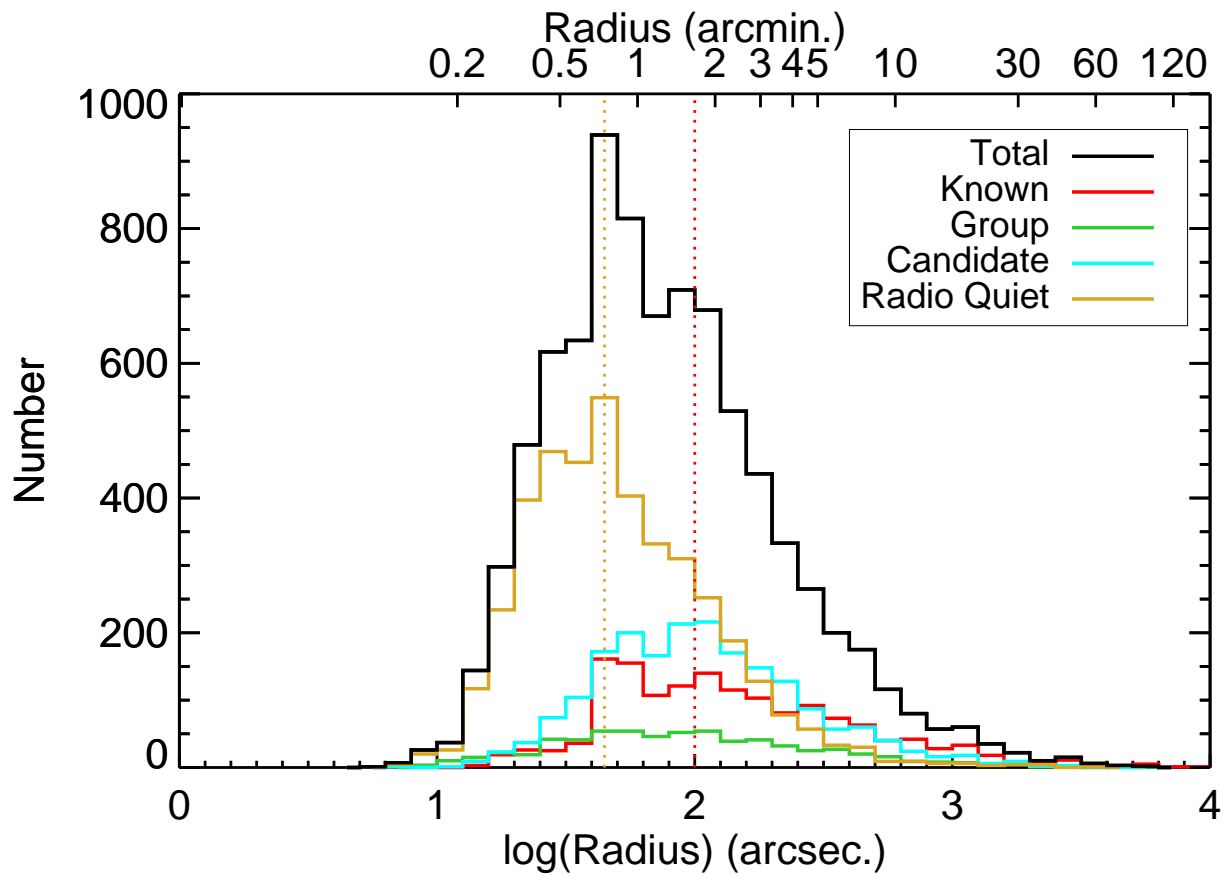}
\caption{Angular size distribution for WISE HII region catalog
  sources.  The known, candidate, and group samples share the same
  distribution.  Radio quiet sources are significantly smaller.  The
  median values for the known and radio quiet samples are shown with
  veritcal dotted lines ($200\arcsec$ and $165\arcsec$,
  respectively). The large difference suggests that the radio quiet
  sample may be a different population of objects.}
\label{fig:sizes}
\end{centering}
\end{figure}

The angular radii of the WISE \hii\ regions range from $6\arcsec$ to
$1\fdg6$.  As described previously, these sizes are visually
determined, and approximately contain the MIR emission of each source.
As seen in Figure~\ref{fig:sizes}, the known and candidate samples
share the same distribution, both with an average angular radius of
$100\arcsec$ (average computed from the log of the radius), and a
  median value approximately the same.  The radio quiet sources have
an average angular radius of $52\arcsec$, and a median value of
  $\sim30\arcsec$.  This hints at a fundamental difference between
the two samples. In Section~\ref{sec:radio-quiet} we argue that this
size difference is a result of the young average age of the radio
quiet sources.  For comparison, the ``large'' MWP bubbles
  have an average angular radius of $\sim150\arcsec$ and a median of
  $\sim130\arcsec$.  The small MWP bubbles average $\sim30\arcsec$,
  with a median also of $\sim30\arcsec$.  Therefore, the angular size
  distribution of the radio-quiet sample is most similar to that of
  the ``small'' MWP bubbles. The angular size distributions of the
  other samples are considerably larger than that of the ``large'' MWP
  bubbles.

\subsection{An Evolved Stellar Shell Seen in Absorption?}
\begin{figure}[!ht]
\begin{centering}
\large
\textsf{
\setlength{\tabcolsep}{.1em}
\begin{tabular}{ccccccc}
4.5\,\micron & 5.8\,\micron & 8.0\,\micron & 12\,\micron & 22\,\micron & 24\,\micron \\
\includegraphics[width=0.9in]{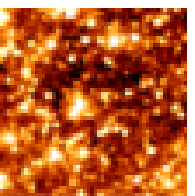} &
\includegraphics[width=0.9in]{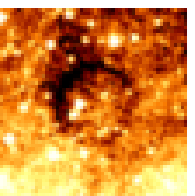} &
\includegraphics[width=0.9in]{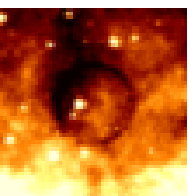} &
\includegraphics[width=0.9in]{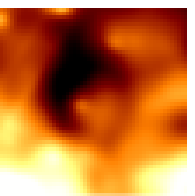} & 
\includegraphics[width=0.9in]{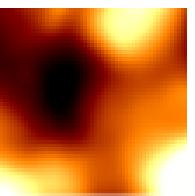} &
\includegraphics[width=0.9in]{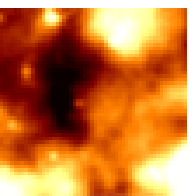} \\
\end{tabular}
}
\caption{The MIR absorption object G030.143+00.228.  Images are
  $2\arcmin$ on a side.}
\label{fig:pn_abs}
\end{centering}
\end{figure}

We detect a bubble that displays infrared absorption, at $\lb\ =
30.143\arcdeg, +0.228\arcdeg$ (Figure~\ref{fig:pn_abs}).  This object
is not in the WISE Catalog of Galactic \hii\ regions; we mention it
here because it is the only such bubble we identified in the over 5000
square degrees examined.  Morphologically, it is similar to the
MIPSGAL bubbles of \citet{mizuno10} in that it is is almost perfectly
circular.  This morphology is in contrast to the less-regular geometry
of most \hii\ regions.  Many of the MIPSGAL bubbles have a central
object but there is nothing detected at MIR wavelengths in the center
of this bubble.  There is also no detected radio continuum emission,
no detection of the object in absorption at wavelengths less than
4.5\,\micron, and no detection in {\it Herschel} Hi-Gal data at
70\,\microns\ \citep{molinari10}.  \citet{mizuno10} argue that most
such objects with circular geometry (detected in emission) are ejecta
from evolved stars and PNe.  If this is the case for this object, it
may allow for the spectroscopic study of the ejecta in absorption.
 
%\subsection{A line of star formation}
%We identify an interesting line of star formation extending from
%\lb\ = (118\arcdeg, 2.5\arcdeg) to (123\arcdeg, 4.0\arcdeg)
%(Figure~\ref{fig:G120}).  Along this line there are no known
%\hii\ regions, but 16 candidates.  Of these candidates, seven have
%molecular line velocities, and five of these seven have a velocity
%component within 4\,\kms\ of $-70\,\kms$.  Five of these seven
%sources, however, have multiple velocity components, which makes
%associating them in velocity difficult.  There is nothing conspicuous
%in the \co\ maps of \citet{heyer98a}.  It is at present unclear if
%this line of star formation regions is a real physical structure.

\section{SUMMARY\label{sec:summary}}
We searched WISE mid-infrared (MIR) images and created a catalog of
Galactic \hii\ regions and \hii\ region candidates.  It improves on
previous Galactic \hii\ region catalogs by removing duplicate sources
and contaminants such as planetary nebulae and supernova remnants.
The catalog extends over all Galactic longitudes within $|b| \le
8\arcdeg$.  We have also included five well-known nearby \hii\ regions
that lie outside this latitude range.  The catalog thus contains all
known Galactic \hii\ regions (        1523
 sources).  Here we define
an \hii\ region as ``known'' if it has been observed in radio
recombination line or H$\alpha$ spectroscopic emission, otherwise it
is an \hii\ region candidate.

By correlating this catalog with previous work, we provide distances
to         1459
 sources.  Over 25\% of all WISE catalog sources
can be associated with molecular emission (        2262
 sources),
most of which do no have ionized gas velocities (        1667
 sources).  There are         1986
 \hii\ region
candidates in the catalog that have radio continuum and MIR emission.
These are ideal objects for followup spectroscopic observations that
can confirm their classification.  Our group's previous RRL studies
show that $\sim95\%$ of such targets are true \hii\ regions.  There
are also         4124
 radio quiet sources that have the
characteristic MIR emission of \hii\ regions, but do not show radio
continuum emission at the sensitivity limits of existing surveys.  The
majority of these are likely in the earliest phases of \hii\ region
evolution, but a significat population are probably real \hii\ regions
with weak, extended radio continuum emission too faint for detection.

The Galactic distribution of the WISE catalog sources is similar to
that of other tracers of star formation, such as \co.  The one
exception is in the Galactic center where there is a deficiency of
sources compared to the amount of molecular material. There is a
slight north-south asymmetry such that $\sim45\%$ of all sources are
located in the first Galactic quadrant and $\sim40\%$ are in the
fourth. We find that $\sim61\%$ of all known Galactic \hii\ regions
are in the first quadrant, due to the larger number of spectroscopic
surveys in this part of the Galaxy. The latitude distribution of WISE
catalog sources is more confined to the Galactic plane compared to
that of CO.

\appendix
\section{THE WISE CATALOG OF GALACTIC H\,{\small \bf II} REGIONS WEB SITE}
We constructed a web site to give others access to the catalog
data\footnote{http://astro.phys.wvu.edu/wise}.  This site has data
from Tables 2 to 6, and provides a visual interface for easily viewing
the catalog contents.  Users can filter the data in myriad ways and
export the output as a machine-readable file.  We will continue to
extend this site as we learn more about the WISE catalog sources.

%%%%%%%%%%%%%%%%%%%%%%%%%%%%%%%%%%%%%%%%%%%%%%%%%%
\begin{acknowledgments}
\nraoblurb.  This research made possible by NASA Astrophysics Data
Processing award number NNX12AI59GADAP to LDA.  Special thanks to the
undergraduate researchers at West Virginia University who helped with
the initial search of the WISE data: Matthew Austin, Meg Barnes,
Stephen Cummings, Kenneth Heitmeyer, April Liska, Eric Newlon, and
especially Alexis Johnson and Devin Williams.  We thank Tom Dame
for help compiling the maser parallax measurements.  This research has
made use of data from the ATLASGAL, BGPS, CGPS, CORNISH, HOPS, MALT90,
MAGPIS, NVSS, SGPS, SUMSS, and VGPS surveys.  The ATLASGAL project is
a collaboration between the Max-Planck-Gesellschaft, the European
Southern Observatory (ESO) and the Universidad de Chile.  This paper
made use of information from the Red MSX Source survey database at
www.ast.leeds.ac.uk/RMS which was constructed with support from the
Science and Technology Facilities Council of the UK.  This research
made use of Montage, funded by the National Aeronautics and Space
Administration's Earth Science Technology Office, Computation
Technologies Project, under Cooperative Agreement Number NCC5-626
between NASA and the California Institute of Technology. Montage is
maintained by the NASA/IPAC Infrared Science Archive. This research
has made use of NASA's Astrophysics Data System Bibliographic Services
and the SIMBAD database operated at CDS, Strasbourg, France.

\end{acknowledgments}

\clearpage
%%%%%%%%%%%%%%%%%%%%%%%%%%%%%%%%%%%%%%%%%%%%%%%%%%
\bibliographystyle{apj} % style aa.bst
\bibliography{ref.bib} % your references Yourfile.bib

\begin{thebibliography}{120}
\expandafter\ifx\csname natexlab\endcsname\relax\def\natexlab#1{#1}\fi

\bibitem[{{Aguirre} {et~al.}(2011){Aguirre}, {Ginsburg}, {Dunham}, {Drosback},
  {Bally}, {Battersby}, {Bradley}, {Cyganowski}, {Dowell}, {Evans}, {Glenn},
  {Harvey}, {Rosolowsky}, {Stringfellow}, {Walawender}, \&
  {Williams}}]{aguirre11}
{Aguirre}, J.~E., {et~al.} 2011, \apjs, 192, 4

\bibitem[{{Anderson}(2010)}]{anderson_thesis}
{Anderson}, L.~D. 2010, PhD thesis, Boston University

\bibitem[{{Anderson} \& {Bania}(2009)}]{anderson09a}
{Anderson}, L.~D., \& {Bania}, T.~M. 2009, \apj, 690, 706

\bibitem[{{Anderson} {et~al.}(2011){Anderson}, {Bania}, {Balser}, \&
  {Rood}}]{anderson11}
{Anderson}, L.~D., {Bania}, T.~M., {Balser}, D.~S., \& {Rood}, R.~T. 2011,
  \apjs, 194, 32

\bibitem[{{Anderson} {et~al.}(2012{\natexlab{a}}){Anderson}, {Bania}, {Balser},
  \& {Rood}}]{anderson12c}
---. 2012{\natexlab{a}}, \apj, 754, 62

\bibitem[{{Anderson} {et~al.}(2009){Anderson}, {Bania}, {Jackson}, {Clemens},
  {Heyer}, {Simon}, {Shah}, \& {Rathborne}}]{anderson09b}
{Anderson}, L.~D., {Bania}, T.~M., {Jackson}, J.~M., {Clemens}, D.~P., {Heyer},
  M., {Simon}, R., {Shah}, R.~Y., \& {Rathborne}, J.~M. 2009, \apjs, 181, 255

\bibitem[{{Anderson} {et~al.}(2012{\natexlab{b}}){Anderson}, {Zavagno},
  {Barlow}, {Garc{\'{\i}}a-Lario}, \& {Noriega-Crespo}}]{anderson12a}
{Anderson}, L.~D., {Zavagno}, A., {Barlow}, M.~J., {Garc{\'{\i}}a-Lario}, P.,
  \& {Noriega-Crespo}, A. 2012{\natexlab{b}}, \aap, 537, A1

\bibitem[{{Anderson} {et~al.}(2012{\natexlab{c}}){Anderson}, {Zavagno},
  {Deharveng}, {Abergel}, {Motte}, {Andr{\'e}}, {Bernard}, {Bontemps},
  {Hennemann}, {Hill}, {Rod{\'o}n}, {Roussel}, \& {Russeil}}]{anderson12b}
{Anderson}, L.~D., {et~al.} 2012{\natexlab{c}}, \aap, 542, A10

\bibitem[{{Andersson} {et~al.}(2000){Andersson}, {Wannier},
  {Moriarty-Schieven}, \& {Bakker}}]{andersson00}
{Andersson}, B.-G., {Wannier}, P.~G., {Moriarty-Schieven}, G.~H., \& {Bakker},
  E.~J. 2000, \aj, 119, 1325

\bibitem[{{Ando} {et~al.}(2011){Ando}, {Nagayama}, {Omodaka}, {Handa}, {Imai},
  {Nakagawa}, {Nakanishi}, {Honma}, {Kobayashi}, \& {Miyaji}}]{ando11}
{Ando}, K., {et~al.} 2011, \pasj, 63, 45

\bibitem[{{Araya} {et~al.}(2002){Araya}, {Hofner}, {Churchwell}, \&
  {Kurtz}}]{araya02}
{Araya}, E., {Hofner}, P., {Churchwell}, E., \& {Kurtz}, S. 2002, \apjs, 138,
  63

\bibitem[{{Arvidsson} {et~al.}(2009){Arvidsson}, {Kerton}, \&
  {Foster}}]{arvidsson09}
{Arvidsson}, K., {Kerton}, C.~R., \& {Foster}, T. 2009, \apj, 700, 1000

\bibitem[{{Balser} {et~al.}(1997){Balser}, {Bania}, {Rood}, \&
  {Wilson}}]{balser97}
{Balser}, D.~S., {Bania}, T.~M., {Rood}, R.~T., \& {Wilson}, T.~L. 1997, \apj,
  483, 320

\bibitem[{{Bania} {et~al.}(2012){Bania}, {Anderson}, \& {Balser}}]{bania12}
{Bania}, T.~M., {Anderson}, L.~D., \& {Balser}, D.~S. 2012, \apj, 759, 96

\bibitem[{{Bania} {et~al.}(2010){Bania}, {Anderson}, {Balser}, \&
  {Rood}}]{bania10}
{Bania}, T.~M., {Anderson}, L.~D., {Balser}, D.~S., \& {Rood}, R.~T. 2010,
  \apjl, 718, L106

\bibitem[{{Bania} \& {Lockman}(1984)}]{bania84}
{Bania}, T.~M., \& {Lockman}, F.~J. 1984, \apjs, 54, 513

\bibitem[{{Bartkiewicz} {et~al.}(2008){Bartkiewicz}, {Brunthaler}, {Szymczak},
  {van Langevelde}, \& {Reid}}]{bartkiewicz08}
{Bartkiewicz}, A., {Brunthaler}, A., {Szymczak}, M., {van Langevelde}, H.~J.,
  \& {Reid}, M.~J. 2008, \aap, 490, 787

\bibitem[{{Becker} {et~al.}(1994){Becker}, {White}, {Helfand}, \&
  {Zoonematkermani}}]{becker94}
{Becker}, R.~H., {White}, R.~L., {Helfand}, D.~J., \& {Zoonematkermani}, S.
  1994, \apjs, 91, 347

\bibitem[{{Benjamin} {et~al.}(2003){Benjamin}, {Churchwell}, {Babler}, {Bania},
  {Clemens}, {Cohen}, {Dickey}, {Indebetouw}, {Jackson}, {Kobulnicky},
  {Lazarian}, {Marston}, {Mathis}, {Meade}, {Seager}, {Stolovy}, {Watson},
  {Whitney}, {Wolff}, \& {Wolfire}}]{benjamin03}
{Benjamin}, R.~A., {et~al.} 2003, \pasp, 115, 953

\bibitem[{{Beuther} {et~al.}(2012){Beuther}, {Tackenberg}, {Linz}, {Henning},
  {Schuller}, {Wyrowski}, {Schilke}, {Menten}, {Robitaille}, {Walmsley},
  {Bronfman}, {Motte}, {Nguyen-Luong}, \& {Bontemps}}]{beuther12}
{Beuther}, H., {et~al.} 2012, \apj, 747, 43

\bibitem[{{Bock} {et~al.}(1999){Bock}, {Large}, \& {Sadler}}]{bock99}
{Bock}, D., {Large}, M.~I., \& {Sadler}, E.~M. 1999, \aj, 117, 1578

\bibitem[{{Brand}(1986)}]{brand86}
{Brand}, J. 1986, PhD thesis, Leiden Univ., Netherlands.

\bibitem[{{Brogan} {et~al.}(2006){Brogan}, {Gelfand}, {Gaensler}, {Kassim}, \&
  {Lazio}}]{brogan06}
{Brogan}, C.~L., {Gelfand}, J.~D., {Gaensler}, B.~M., {Kassim}, N.~E., \&
  {Lazio}, T.~J.~W. 2006, \apjl, 639, L25

\bibitem[{{Bronfman} {et~al.}(1996){Bronfman}, {Nyman}, \& {May}}]{bronfman96}
{Bronfman}, L., {Nyman}, L.-A., \& {May}, J. 1996, \aaps, 115, 81

\bibitem[{{Brunthaler} {et~al.}(2009){Brunthaler}, {Reid}, {Menten}, {Zheng},
  {Moscadelli}, \& {Xu}}]{brunthaler09}
{Brunthaler}, A., {Reid}, M.~J., {Menten}, K.~M., {Zheng}, X.~W., {Moscadelli},
  L., \& {Xu}, Y. 2009, \apj, 693, 424

\bibitem[{{Brunthaler} {et~al.}(2011){Brunthaler}, {Reid}, {Menten}, {Zheng},
  {Bartkiewicz}, {Choi}, {Dame}, {Hachisuka}, {Immer}, {Moellenbrock},
  {Moscadelli}, {Rygl}, {Sanna}, {Sato}, {Wu}, {Xu}, \& {Zhang}}]{brunthaler11}
{Brunthaler}, A., {et~al.} 2011, Astronomische Nachrichten, 332, 461

\bibitem[{{Carey} {et~al.}(2009){Carey}, {Noriega-Crespo}, {Mizuno}, {Shenoy},
  {Paladini}, {Kraemer}, {Price}, {Flagey}, {Ryan}, {Ingalls}, {Kuchar},
  {Pinheiro Gon{\c c}alves}, {Indebetouw}, {Billot}, {Marleau}, {Padgett},
  {Rebull}, {Bressert}, {Ali}, {Molinari}, {Martin}, {Berriman}, {Boulanger},
  {Latter}, {Miville-Deschenes}, {Shipman}, \& {Testi}}]{carey09}
{Carey}, S.~J., {et~al.} 2009, \pasp, 121, 76

\bibitem[{{Caswell} \& {Haynes}(1987)}]{caswell87}
{Caswell}, J.~L., \& {Haynes}, R.~F. 1987, \aap, 171, 261

\bibitem[{{Churchwell} {et~al.}(2009){Churchwell}, {Babler}, {Meade},
  {Whitney}, {Benjamin}, {Indebetouw}, {Cyganowski}, {Robitaille}, {Povich},
  {Watson}, \& {Bracker}}]{churchwell09}
{Churchwell}, E., {et~al.} 2009, \pasp, 121, 213

\bibitem[{{Clark} {et~al.}(2003){Clark}, {Egan}, {Crowther}, {Mizuno},
  {Larionov}, \& {Arkharov}}]{clark03}
{Clark}, J.~S., {Egan}, M.~P., {Crowther}, P.~A., {Mizuno}, D.~R., {Larionov},
  V.~M., \& {Arkharov}, A. 2003, \aap, 412, 185

\bibitem[{{Condon} {et~al.}(1998){Condon}, {Cotton}, {Greisen}, {Yin},
  {Perley}, {Taylor}, \& {Broderick}}]{condon98}
{Condon}, J.~J., {Cotton}, W.~D., {Greisen}, E.~W., {Yin}, Q.~F., {Perley},
  R.~A., {Taylor}, G.~B., \& {Broderick}, J.~J. 1998, \aj, 115, 1693

\bibitem[{{Contreras} {et~al.}(2013){Contreras}, {Schuller}, {Urquhart},
  {Csengeri}, {Wyrowski}, {Beuther}, {Bontemps}, {Bronfman}, {Henning},
  {Menten}, {Schilke}, {Walmsley}, {Wienen}, {Tackenberg}, \&
  {Linz}}]{contreras13}
{Contreras}, Y., {et~al.} 2013, \aap, 549, A45

\bibitem[{{Dame} {et~al.}(2001){Dame}, {Hartmann}, \& {Thaddeus}}]{dame01}
{Dame}, T.~M., {Hartmann}, D., \& {Thaddeus}, P. 2001, \apj, 547, 792

\bibitem[{{Deharveng} {et~al.}(2003){Deharveng}, {Lefloch}, {Zavagno},
  {Caplan}, {Whitworth}, {Nadeau}, \& {Mart{\'{\i}}n}}]{deharveng03}
{Deharveng}, L., {Lefloch}, B., {Zavagno}, A., {Caplan}, J., {Whitworth},
  A.~P., {Nadeau}, D., \& {Mart{\'{\i}}n}, S. 2003, \aap, 408, L25

\bibitem[{{Deharveng} {et~al.}(2010){Deharveng}, {Schuller}, {Anderson},
  {Zavagno}, {Wyrowski}, {Menten}, {Bronfman}, {Testi}, {Walmsley}, \&
  {Wienen}}]{deharveng10}
{Deharveng}, L., {et~al.} 2010, \aap, 523, A6+

\bibitem[{{Downes} {et~al.}(1980){Downes}, {Wilson}, {Bieging}, \&
  {Wink}}]{downes80}
{Downes}, D., {Wilson}, T.~L., {Bieging}, J., \& {Wink}, J. 1980, \aaps, 40,
  379

\bibitem[{{Dunham} {et~al.}(2011){Dunham}, {Rosolowsky}, {Evans}, {Cyganowski},
  \& {Urquhart}}]{dunham11}
{Dunham}, M.~K., {Rosolowsky}, E., {Evans}, II, N.~J., {Cyganowski}, C., \&
  {Urquhart}, J.~S. 2011, \apj, 741, 110

\bibitem[{{Fazio} {et~al.}(2004){Fazio}, {Hora}, {Allen}, {Ashby}, {Barmby},
  {Deutsch}, {Huang}, {Kleiner}, {Marengo}, {Megeath}, {Melnick}, {Pahre},
  {Patten}, {Polizotti}, {Smith}, {Taylor}, {Wang}, {Willner}, {Hoffmann},
  {Pipher}, {Forrest}, {McMurty}, {McCreight}, {McKelvey}, {McMurray}, {Koch},
  {Moseley}, {Arendt}, {Mentzell}, {Marx}, {Losch}, {Mayman}, {Eichhorn},
  {Krebs}, {Jhabvala}, {Gezari}, {Fixsen}, {Flores}, {Shakoorzadeh}, {Jungo},
  {Hakun}, {Workman}, {Karpati}, {Kichak}, {Whitley}, {Mann}, {Tollestrup},
  {Eisenhardt}, {Stern}, {Gorjian}, {Bhattacharya}, {Carey}, {Nelson},
  {Glaccum}, {Lacy}, {Lowrance}, {Laine}, {Reach}, {Stauffer}, {Surace},
  {Wilson}, {Wright}, {Hoffman}, {Domingo}, \& {Cohen}}]{fazio04}
{Fazio}, G.~G., {et~al.} 2004, \apjs, 154, 10

\bibitem[{{Fich} {et~al.}(1990){Fich}, {Dahl}, \& {Treffers}}]{fich90}
{Fich}, M., {Dahl}, G.~P., \& {Treffers}, R.~R. 1990, \aj, 99, 622

\bibitem[{{Flaherty} {et~al.}(2007){Flaherty}, {Pipher}, {Megeath}, {Winston},
  {Gutermuth}, {Muzerolle}, {Allen}, \& {Fazio}}]{flaherty07}
{Flaherty}, K.~M., {Pipher}, J.~L., {Megeath}, S.~T., {Winston}, E.~M.,
  {Gutermuth}, R.~A., {Muzerolle}, J., {Allen}, L.~E., \& {Fazio}, G.~G. 2007,
  \apj, 663, 1069

\bibitem[{{Foster} {et~al.}(2011){Foster}, {Jackson}, {Barnes}, {Barris},
  {Brooks}, {Cunningham}, {Finn}, {Fuller}, {Longmore}, {Mascoop}, {Peretto},
  {Rathborne}, {Sanhueza}, {Schuller}, \& {Wyrowski}}]{foster11}
{Foster}, J.~B., {et~al.} 2011, \apjs, 197, 25

\bibitem[{{Gaensler} {et~al.}(1999){Gaensler}, {Gotthelf}, \&
  {Vasisht}}]{gaensler99}
{Gaensler}, B.~M., {Gotthelf}, E.~V., \& {Vasisht}, G. 1999, \apjl, 526, L37

\bibitem[{{Garay} {et~al.}(1989){Garay}, {Gathier}, \& {Rodriguez}}]{garay89}
{Garay}, G., {Gathier}, R., \& {Rodriguez}, L.~F. 1989, \aap, 215, 101

\bibitem[{{G{\'o}mez}(2006)}]{gomez06}
{G{\'o}mez}, G.~C. 2006, \aj, 132, 2376

\bibitem[{{Gregory} {et~al.}(1996){Gregory}, {Scott}, {Douglas}, \&
  {Condon}}]{gregory96}
{Gregory}, P.~C., {Scott}, W.~K., {Douglas}, K., \& {Condon}, J.~J. 1996,
  \apjs, 103, 427

\bibitem[{{Hachisuka} {et~al.}(2009){Hachisuka}, {Brunthaler}, {Menten},
  {Reid}, {Hagiwara}, \& {Mochizuki}}]{hachisuka09}
{Hachisuka}, K., {Brunthaler}, A., {Menten}, K.~M., {Reid}, M.~J., {Hagiwara},
  Y., \& {Mochizuki}, N. 2009, \apj, 696, 1981

\bibitem[{{Helfand} {et~al.}(2006){Helfand}, {Becker}, {White}, {Fallon}, \&
  {Tuttle}}]{helfand06}
{Helfand}, D.~J., {Becker}, R.~H., {White}, R.~L., {Fallon}, A., \& {Tuttle},
  S. 2006, \aj, 131, 2525

\bibitem[{{Hirota} {et~al.}(2008{\natexlab{a}}){Hirota}, {Bushimata}, {Choi},
  {Honma}, {Imai}, {Iwadate}, {Jike}, {Kameya}, {Kamohara}, {Kan-Ya},
  {Kawaguchi}, {Kijima}, {Kobayashi}, {Kuji}, {Kurayama}, {Manabe}, {Miyaji},
  {Nagayama}, {Nakagawa}, {Oh}, {Omodaka}, {Oyama}, {Sakai}, {Sasao}, {Sato},
  {Shibata}, {Tamura}, \& {Yamashita}}]{hirota08}
{Hirota}, T., {et~al.} 2008{\natexlab{a}}, \pasj, 60, 37

\bibitem[{{Hirota} {et~al.}(2008{\natexlab{b}}){Hirota}, {Bushimata}, {Choi},
  {Honma}, {Imai}, {Iwadate}, {Jike}, {Kameya}, {Kamohara}, {Kan-Ya},
  {Kawaguchi}, {Kijima}, {Kobayashi}, {Kuji}, {Kurayama}, {Manabe}, {Miyaji},
  {Nagayama}, {Nakagawa}, {Oh}, {Omodaka}, {Oyama}, {Sakai}, {Sasao}, {Sato},
  {Shibata}, {Tamura}, \& {Yamashita}}]{hirota08a}
---. 2008{\natexlab{b}}, \pasj, 60, 37

\bibitem[{{Hirota} {et~al.}(2008{\natexlab{c}}){Hirota}, {Ando}, {Bushimata},
  {Choi}, {Honma}, {Imai}, {Iwadate}, {Jike}, {Kameno}, {Kameya}, {Kamohara},
  {Kan-Ya}, {Kawaguchi}, {Kijima}, {Kim}, {Kobayashi}, {Kuji}, {Kurayama},
  {Manabe}, {Matsui}, {Matsumoto}, {Miyaji}, {Miyazaki}, {Nagayama},
  {Nakagawa}, {Namikawa}, {Nyu}, {Oh}, {Omodaka}, {Oyama}, {Sakai}, {Sasao},
  {Sato}, {Sato}, {Shibata}, {Tamura}, {Ueda}, \& {Yamashita}}]{hirota08b}
---. 2008{\natexlab{c}}, \pasj, 60, 961

\bibitem[{{Hoare} {et~al.}(2012){Hoare}, {Purcell}, {Churchwell}, {Diamond},
  {Cotton}, {Chandler}, {Smethurst}, {Kurtz}, {Mundy}, {Dougherty}, {Fender},
  {Fuller}, {Jackson}, {Garrington}, {Gledhill}, {Goldsmith}, {Lumsden},
  {Mart{\'{\i}}}, {Moore}, {Muxlow}, {Oudmaijer}, {Pandian}, {Paredes},
  {Shepherd}, {Spencer}, {Thompson}, {Umana}, {Urquhart}, \&
  {Zijlstra}}]{hoare12}
{Hoare}, M.~G., {et~al.} 2012, \pasp, 124, 939

\bibitem[{{Honma} {et~al.}(2007){Honma}, {Bushimata}, {Choi}, {Hirota}, {Imai},
  {Iwadate}, {Jike}, {Kameya}, {Kamohara}, {Kan-Ya}, {Kawaguchi}, {Kijima},
  {Kobayashi}, {Kuji}, {Kurayama}, {Manabe}, {Miyaji}, {Nagayama}, {Nakagawa},
  {Oh}, {Omodaka}, {Oyama}, {Sakai}, {Sato}, {Sasao}, {Shibata}, {Shintani},
  {Suda}, {Tamura}, {Tsushima}, \& {Yamashita Kazuyoshi}}]{honma07}
{Honma}, M., {et~al.} 2007, \pasj, 59, 889

\bibitem[{{Immer} {et~al.}(2013){Immer}, {Reid}, {Menten}, {Brunthaler}, \&
  {Dame}}]{immer13}
{Immer}, K., {Reid}, M.~J., {Menten}, K.~M., {Brunthaler}, A., \& {Dame}, T.~M.
  2013, \aap, 553, A117

\bibitem[{{Jackson} {et~al.}(2006){Jackson}, {Rathborne}, {Shah}, {Simon},
  {Bania}, {Clemens}, {Chambers}, {Johnson}, {Dormody}, {Lavoie}, \&
  {Heyer}}]{jackson06}
{Jackson}, J.~M., {et~al.} 2006, \apjs, 163, 145

\bibitem[{{Jones} \& {Dickey}(2012)}]{jones12}
{Jones}, C., \& {Dickey}, J.~M. 2012, \apj, 753, 62

\bibitem[{{Kobulnicky} {et~al.}(2013){Kobulnicky}, {Babler}, {Alexander},
  {Meade}, {Whitney}, \& {Churchwell}}]{kobulnicky13}
{Kobulnicky}, H.~A., {Babler}, B.~L., {Alexander}, M.~J., {Meade}, M.~R.,
  {Whitney}, B.~A., \& {Churchwell}, E.~B. 2013, \apjs, 207, 9

\bibitem[{{Kurayama} {et~al.}(2011){Kurayama}, {Nakagawa}, {Sawada-Satoh},
  {Sato}, {Honma}, {Sunada}, {Hirota}, \& {Imai}}]{kurayama11}
{Kurayama}, T., {Nakagawa}, A., {Sawada-Satoh}, S., {Sato}, K., {Honma}, M.,
  {Sunada}, K., {Hirota}, T., \& {Imai}, H. 2011, \pasj, 63, 513

\bibitem[{{Kurtz} {et~al.}(1994){Kurtz}, {Churchwell}, \& {Wood}}]{kurtz94}
{Kurtz}, S., {Churchwell}, E., \& {Wood}, D.~O.~S. 1994, \apjs, 91, 659

\bibitem[{{Lockman}(1989)}]{lockman89}
{Lockman}, F.~J. 1989, \apjs, 71, 469

\bibitem[{{Lockman} {et~al.}(1996){Lockman}, {Pisano}, \& {Howard}}]{lockman96}
{Lockman}, F.~J., {Pisano}, D.~J., \& {Howard}, G.~J. 1996, \apj, 472, 173

\bibitem[{{Longmore} {et~al.}(2013){Longmore}, {Bally}, {Testi}, {Purcell},
  {Walsh}, {Bressert}, {Pestalozzi}, {Molinari}, {Ott}, {Cortese}, {Battersby},
  {Murray}, {Lee}, {Kruijssen}, {Schisano}, \& {Elia}}]{longmore13}
{Longmore}, S.~N., {et~al.} 2013, \mnras, 429, 987

\bibitem[{{McClure-Griffiths} {et~al.}(2005){McClure-Griffiths}, {Dickey},
  {Gaensler}, {Green}, {Haverkorn}, \& {Strasser}}]{mcclure05}
{McClure-Griffiths}, N.~M., {Dickey}, J.~M., {Gaensler}, B.~M., {Green}, A.~J.,
  {Haverkorn}, M., \& {Strasser}, S. 2005, \apjs, 158, 178

\bibitem[{{Menten} {et~al.}(2007){Menten}, {Reid}, {Forbrich}, \&
  {Brunthaler}}]{menten07}
{Menten}, K.~M., {Reid}, M.~J., {Forbrich}, J., \& {Brunthaler}, A. 2007, \aap,
  474, 515

\bibitem[{{Mizuno} {et~al.}(2010){Mizuno}, {Kraemer}, {Flagey}, {Billot},
  {Shenoy}, {Paladini}, {Ryan}, {Noriega-Crespo}, \& {Carey}}]{mizuno10}
{Mizuno}, D.~R., {et~al.} 2010, \aj, 139, 1542

\bibitem[{{Moellenbrock} {et~al.}(2009){Moellenbrock}, {Claussen}, \&
  {Goss}}]{moellenbrock09}
{Moellenbrock}, G.~A., {Claussen}, M.~J., \& {Goss}, W.~M. 2009, \apj, 694, 192

\bibitem[{{Molinari} {et~al.}(2010){Molinari}, {Swinyard}, {Bally}, {Barlow},
  {Bernard}, {Martin}, {Moore}, {Noriega-Crespo}, {Plume}, {Testi}, {Zavagno},
  {Abergel}, {Ali}, {Anderson}, {Andr{\'e}}, {Baluteau}, {Battersby},
  {Beltr{\'a}n}, {Benedettini}, {Billot}, {Blommaert}, {Bontemps}, {Boulanger},
  {Brand}, {Brunt}, {Burton}, {Calzoletti}, {Carey}, {Caselli}, {Cesaroni},
  {Cernicharo}, {Chakrabarti}, {Chrysostomou}, {Cohen}, {Compiegne}, {de
  Bernardis}, {de Gasperis}, {di Giorgio}, {Elia}, {Faustini}, {Flagey},
  {Fukui}, {Fuller}, {Ganga}, {Garcia-Lario}, {Glenn}, {Goldsmith}, {Griffin},
  {Hoare}, {Huang}, {Ikhenaode}, {Joblin}, {Joncas}, {Juvela}, {Kirk},
  {Lagache}, {Li}, {Lim}, {Lord}, {Marengo}, {Marshall}, {Masi}, {Massi},
  {Matsuura}, {Minier}, {Miville-Desch{\^e}nes}, {Montier}, {Morgan}, {Motte},
  {Mottram}, {M{\"u}ller}, {Natoli}, {Neves}, {Olmi}, {Paladini}, {Paradis},
  {Parsons}, {Peretto}, {Pestalozzi}, {Pezzuto}, {Piacentini}, {Piazzo},
  {Polychroni}, {Pomar{\`e}s}, {Popescu}, {Reach}, {Ristorcelli}, {Robitaille},
  {Robitaille}, {Rod{\'o}n}, {Roy}, {Royer}, {Russeil}, {Saraceno}, {Sauvage},
  {Schilke}, {Schisano}, {Schneider}, {Schuller}, {Schulz}, {Sibthorpe},
  {Smith}, {Smith}, {Spinoglio}, {Stamatellos}, {Strafella}, {Stringfellow},
  {Sturm}, {Taylor}, {Thompson}, {Traficante}, {Tuffs}, {Umana}, {Valenziano},
  {Vavrek}, {Veneziani}, {Viti}, {Waelkens}, {Ward-Thompson}, {White},
  {Wilcock}, {Wyrowski}, {Yorke}, \& {Zhang}}]{molinari10}
{Molinari}, S., {et~al.} 2010, \aap, 518, L100+

\bibitem[{{Moscadelli} {et~al.}(2009){Moscadelli}, {Reid}, {Menten},
  {Brunthaler}, {Zheng}, \& {Xu}}]{moscadelli09}
{Moscadelli}, L., {Reid}, M.~J., {Menten}, K.~M., {Brunthaler}, A., {Zheng},
  X.~W., \& {Xu}, Y. 2009, \apj, 693, 406

\bibitem[{{Niinuma} {et~al.}(2011){Niinuma}, {Nagayama}, {Hirota}, {Honma},
  {Motogi}, {Nakagawa}, {Kurayama}, {Kan-Ya}, {Kawaguchi}, {Kobayashi}, \&
  {Ueno}}]{niinuma11}
{Niinuma}, K., {et~al.} 2011, \pasj, 63, 9

\bibitem[{{Oh} {et~al.}(2010){Oh}, {Kobayashi}, {Honma}, {Hirota}, {Sato}, \&
  {Ueno}}]{oh10}
{Oh}, C.~S., {Kobayashi}, H., {Honma}, M., {Hirota}, T., {Sato}, K., \& {Ueno},
  Y. 2010, \pasj, 62, 101

\bibitem[{{Paladini} {et~al.}(2003){Paladini}, {Burigana}, {Davies}, {Maino},
  {Bersanelli}, {Cappellini}, {Platania}, \& {Smoot}}]{paladini03}
{Paladini}, R., {Burigana}, C., {Davies}, R.~D., {Maino}, D., {Bersanelli}, M.,
  {Cappellini}, B., {Platania}, P., \& {Smoot}, G. 2003, \aap, 397, 213

\bibitem[{{Purcell} {et~al.}(2012){Purcell}, {Longmore}, {Walsh}, {Whiting},
  {Breen}, {Britton}, {Brooks}, {Burton}, {Cunningham}, {Green},
  {Harvey-Smith}, {Hindson}, {Hoare}, {Indermuehle}, {Jones}, {Lo}, {Lowe},
  {Phillips}, {Thompson}, {Urquhart}, {Voronkov}, \& {White}}]{purcell12}
{Purcell}, C.~R., {et~al.} 2012, \mnras, 426, 1972

\bibitem[{{Purcell} {et~al.}(2013){Purcell}, {Hoare}, {Cotton}, {Lumsden},
  {Urquhart}, {Chandler}, {Churchwell}, {Diamond}, {Dougherty}, {Fender},
  {Fuller}, {Garrington}, {Gledhill}, {Goldsmith}, {Hindson}, {Jackson},
  {Kurtz}, {Mart{\'{\i}}}, {Moore}, {Mundy}, {Muxlow}, {Oudmaijer}, {Pandian},
  {Paredes}, {Shepherd}, {Smethurst}, {Spencer}, {Thompson}, {Umana}, \&
  {Zijlstra}}]{purcell13}
---. 2013, \apjs, 205, 1

\bibitem[{{Reid} {et~al.}(2009{\natexlab{a}}){Reid}, {Menten}, {Brunthaler},
  {Zheng}, {Moscadelli}, \& {Xu}}]{reid09a}
{Reid}, M.~J., {Menten}, K.~M., {Brunthaler}, A., {Zheng}, X.~W., {Moscadelli},
  L., \& {Xu}, Y. 2009{\natexlab{a}}, \apj, 693, 397

\bibitem[{{Reid} {et~al.}(2009{\natexlab{b}}){Reid}, {Menten}, {Zheng},
  {Brunthaler}, {Moscadelli}, {Xu}, {Zhang}, {Sato}, {Honma}, {Hirota},
  {Hachisuka}, {Choi}, {Moellenbrock}, \& {Bartkiewicz}}]{reid09b}
{Reid}, M.~J., {et~al.} 2009{\natexlab{b}}, \apj, 700, 137

\bibitem[{{Rieke} \& {Lebofsky}(1985)}]{rieke85}
{Rieke}, G.~H., \& {Lebofsky}, M.~J. 1985, \apj, 288, 618

\bibitem[{{Rieke} {et~al.}(2004){Rieke}, {Young}, {Engelbracht}, {Kelly},
  {Low}, {Haller}, {Beeman}, {Gordon}, {Stansberry}, {Misselt}, {Cadien},
  {Morrison}, {Rivlis}, {Latter}, {Noriega-Crespo}, {Padgett}, {Stapelfeldt},
  {Hines}, {Egami}, {Muzerolle}, {Alonso-Herrero}, {Blaylock}, {Dole}, {Hinz},
  {Le Floc'h}, {Papovich}, {P{\'e}rez-Gonz{\'a}lez}, {Smith}, {Su}, {Bennett},
  {Frayer}, {Henderson}, {Lu}, {Masci}, {Pesenson}, {Rebull}, {Rho}, {Keene},
  {Stolovy}, {Wachter}, {Wheaton}, {Werner}, \& {Richards}}]{rieke04}
{Rieke}, G.~H., {et~al.} 2004, \apjs, 154, 25

\bibitem[{{Roman-Duval} {et~al.}(2009){Roman-Duval}, {Jackson}, {Heyer},
  {Johnson}, {Rathborne}, {Shah}, \& {Simon}}]{roman-duval09}
{Roman-Duval}, J., {Jackson}, J.~M., {Heyer}, M., {Johnson}, A., {Rathborne},
  J., {Shah}, R., \& {Simon}, R. 2009, ArXiv e-prints

\bibitem[{{Rosolowsky} {et~al.}(2010){Rosolowsky}, {Dunham}, {Ginsburg},
  {Bradley}, {Aguirre}, {Bally}, {Battersby}, {Cyganowski}, {Dowell},
  {Drosback}, {Evans}, {Glenn}, {Harvey}, {Stringfellow}, {Walawender}, \&
  {Williams}}]{rosolowsky10}
{Rosolowsky}, E., {et~al.} 2010, \apjs, 188, 123

\bibitem[{{Rubin}(1968)}]{rubin68}
{Rubin}, R.~H. 1968, \apj, 154, 391

\bibitem[{{Russeil} {et~al.}(2007){Russeil}, {Adami}, \&
  {Georgelin}}]{russeil07}
{Russeil}, D., {Adami}, C., \& {Georgelin}, Y.~M. 2007, \aap, 470, 161

\bibitem[{{Rygl} {et~al.}(2010){Rygl}, {Brunthaler}, {Reid}, {Menten}, {van
  Langevelde}, \& {Xu}}]{rygl10}
{Rygl}, K.~L.~J., {Brunthaler}, A., {Reid}, M.~J., {Menten}, K.~M., {van
  Langevelde}, H.~J., \& {Xu}, Y. 2010, \aap, 511, A2

\bibitem[{{Rygl} {et~al.}(2012){Rygl}, {Brunthaler}, {Sanna}, {Menten}, {Reid},
  {van Langevelde}, {Honma}, {Torstensson}, \& {Fujisawa}}]{rygl12}
{Rygl}, K.~L.~J., {et~al.} 2012, \aap, 539, A79

\bibitem[{{Sanchez-Monge} {et~al.}(2013){Sanchez-Monge}, {Palau}, {Fontani},
  {Busquet}, {Juarez}, {Estalella}, {Tan}, {Sepulveda}, {Ho}, {Zhang}, \&
  {Kurtz}}]{sanchez-monge13}
{Sanchez-Monge}, A., {et~al.} 2013, ArXiv e-prints

\bibitem[{{Sanna} {et~al.}(2012){Sanna}, {Reid}, {Dame}, {Menten},
  {Brunthaler}, {Moscadelli}, {Zheng}, \& {Xu}}]{sanna12}
{Sanna}, A., {Reid}, M.~J., {Dame}, T.~M., {Menten}, K.~M., {Brunthaler}, A.,
  {Moscadelli}, L., {Zheng}, X.~W., \& {Xu}, Y. 2012, \apj, 745, 82

\bibitem[{{Sanna} {et~al.}(2009){Sanna}, {Reid}, {Moscadelli}, {Dame},
  {Menten}, {Brunthaler}, {Zheng}, \& {Xu}}]{sanna09}
{Sanna}, A., {Reid}, M.~J., {Moscadelli}, L., {Dame}, T.~M., {Menten}, K.~M.,
  {Brunthaler}, A., {Zheng}, X.~W., \& {Xu}, Y. 2009, \apj, 706, 464

\bibitem[{{Sato} {et~al.}(2010{\natexlab{a}}){Sato}, {Hirota}, {Reid}, {Honma},
  {Kobayashi}, {Iwadate}, {Miyaji}, \& {Shibata}}]{sato10a}
{Sato}, M., {Hirota}, T., {Reid}, M.~J., {Honma}, M., {Kobayashi}, H.,
  {Iwadate}, K., {Miyaji}, T., \& {Shibata}, K.~M. 2010{\natexlab{a}}, \pasj,
  62, 287

\bibitem[{{Sato} {et~al.}(2010{\natexlab{b}}){Sato}, {Reid}, {Brunthaler}, \&
  {Menten}}]{sato10b}
{Sato}, M., {Reid}, M.~J., {Brunthaler}, A., \& {Menten}, K.~M.
  2010{\natexlab{b}}, \apj, 720, 1055

\bibitem[{{Sato} {et~al.}(2008){Sato}, {Hirota}, {Honma}, {Kobayashi}, {Sasao},
  {Bushimata}, {Choi}, {Imai}, {Iwadate}, {Jike}, {Kameno}, {Kameya},
  {Kamohara}, {Kan-Ya}, {Kawaguchi}, {Kim}, {Kuji}, {Kurayama}, {Manabe},
  {Matsui}, {Matsumoto}, {Miyaji}, {Nagayama}, {Nakagawa}, {Nakamura}, {Oh},
  {Omodaka}, {Oyama}, {Sakai}, {Sato}, {Shibata}, {Tamura}, \&
  {Yamashita}}]{sato08}
{Sato}, M., {et~al.} 2008, \pasj, 60, 975

\bibitem[{{Schlingman} {et~al.}(2011){Schlingman}, {Shirley}, {Schenk},
  {Rosolowsky}, {Bally}, {Battersby}, {Dunham}, {Ellsworth-Bowers}, {Evans},
  {Ginsburg}, \& {Stringfellow}}]{schlingman11}
{Schlingman}, W.~M., {et~al.} 2011, \apjs, 195, 14

\bibitem[{{Schuller} {et~al.}(2009){Schuller}, {Menten}, {Contreras},
  {Wyrowski}, {Schilke}, {Bronfman}, {Henning}, {Walmsley}, {Beuther},
  {Bontemps}, {Cesaroni}, {Deharveng}, {Garay}, {Herpin}, {Lefloch}, {Linz},
  {Mardones}, {Minier}, {Molinari}, {Motte}, {Nyman}, {Reveret}, {Risacher},
  {Russeil}, {Schneider}, {Testi}, {Troost}, {Vasyunina}, {Wienen}, {Zavagno},
  {Kovacs}, {Kreysa}, {Siringo}, \& {Wei{\ss}}}]{schuller09}
{Schuller}, F., {et~al.} 2009, \aap, 504, 415

\bibitem[{{Sewilo} {et~al.}(2004{\natexlab{a}}){Sewilo}, {Churchwell}, {Kurtz},
  {Goss}, \& {Hofner}}]{sewilo04a}
{Sewilo}, M., {Churchwell}, E., {Kurtz}, S., {Goss}, W.~M., \& {Hofner}, P.
  2004{\natexlab{a}}, \apj, 605, 285

\bibitem[{{Sewilo} {et~al.}(2004{\natexlab{b}}){Sewilo}, {Watson}, {Araya},
  {Churchwell}, {Hofner}, \& {Kurtz}}]{sewilo04b}
{Sewilo}, M., {Watson}, C., {Araya}, E., {Churchwell}, E., {Hofner}, P., \&
  {Kurtz}, S. 2004{\natexlab{b}}, \apjs, 154, 553

\bibitem[{{Shiozaki} {et~al.}(2011){Shiozaki}, {Imai}, {Tafoya}, {Omodaka},
  {Hirota}, {Honma}, {Matsui}, \& {Ueno}}]{shiozaki11}
{Shiozaki}, S., {Imai}, H., {Tafoya}, D., {Omodaka}, T., {Hirota}, T., {Honma},
  M., {Matsui}, M., \& {Ueno}, Y. 2011, \pasj, 63, 1219

\bibitem[{{Simpson} {et~al.}(2012){Simpson}, {Povich}, {Kendrew}, {Lintott},
  {Bressert}, {Arvidsson}, {Cyganowski}, {Maddison}, {Schawinski}, {Sherman},
  {Smith}, \& {Wolf-Chase}}]{simpson12}
{Simpson}, R.~J., {et~al.} 2012, ArXiv e-prints

\bibitem[{{Spitzer}(1968)}]{spitzer}
{Spitzer}, L. 1968, {Diffuse matter in space} (New York: Interscience
  Publication, 1968)

\bibitem[{{Sternberg} {et~al.}(2003){Sternberg}, {Hoffmann}, \&
  {Pauldrach}}]{sternberg03}
{Sternberg}, A., {Hoffmann}, T.~L., \& {Pauldrach}, A.~W.~A. 2003, \apj, 599,
  1333

\bibitem[{{Stil} {et~al.}(2006){Stil}, {Taylor}, {Dickey}, {Kavars}, {Martin},
  {Rothwell}, {Boothroyd}, {Lockman}, \& {McClure-Griffiths}}]{stil06}
{Stil}, J.~M., {et~al.} 2006, \aj, 132, 1158

\bibitem[{{Taylor} {et~al.}(2003){Taylor}, {Gibson}, {Peracaula}, {Martin},
  {Landecker}, {Brunt}, {Dewdney}, {Dougherty}, {Gray}, {Higgs}, {Kerton},
  {Knee}, {Kothes}, {Purton}, {Uyaniker}, {Wallace}, {Willis}, \&
  {Durand}}]{taylor03}
{Taylor}, A.~R., {et~al.} 2003, \aj, 125, 3145

\bibitem[{{Tielens}(2008)}]{tielens08}
{Tielens}, A.~G.~G.~M. 2008, \araa, 46, 289

\bibitem[{{Urquhart} {et~al.}(2007{\natexlab{a}}){Urquhart}, {Busfield},
  {Hoare}, {Lumsden}, {Clarke}, {Moore}, {Mottram}, \&
  {Oudmaijer}}]{urquhart07a}
{Urquhart}, J.~S., {Busfield}, A.~L., {Hoare}, M.~G., {Lumsden}, S.~L.,
  {Clarke}, A.~J., {Moore}, T.~J.~T., {Mottram}, J.~C., \& {Oudmaijer}, R.~D.
  2007{\natexlab{a}}, \aap, 461, 11

\bibitem[{{Urquhart} {et~al.}(2008{\natexlab{a}}){Urquhart}, {Hoare},
  {Lumsden}, {Oudmaijer}, \& {Moore}}]{urquhart08a}
{Urquhart}, J.~S., {Hoare}, M.~G., {Lumsden}, S.~L., {Oudmaijer}, R.~D., \&
  {Moore}, T.~J.~T. 2008{\natexlab{a}}, in Astronomical Society of the Pacific
  Conference Series, Vol. 387, Massive Star Formation: Observations Confront
  Theory, ed. {H.~Beuther, H.~Linz, \& T.~Henning}, 381--+

\bibitem[{{Urquhart} {et~al.}(2007{\natexlab{b}}){Urquhart}, {Busfield},
  {Hoare}, {Lumsden}, {Oudmaijer}, {Moore}, {Gibb}, {Purcell}, {Burton}, \&
  {Marechal}}]{urquhart07b}
{Urquhart}, J.~S., {et~al.} 2007{\natexlab{b}}, \aap, 474, 891

\bibitem[{{Urquhart} {et~al.}(2008{\natexlab{b}}){Urquhart}, {Busfield},
  {Hoare}, {Lumsden}, {Oudmaijer}, {Moore}, {Gibb}, {Purcell}, {Burton},
  {Mar{\'e}chal}, {Jiang}, \& {Wang}}]{urquhart08b}
---. 2008{\natexlab{b}}, \aap, 487, 253

\bibitem[{{Urquhart} {et~al.}(2009){Urquhart}, {Hoare}, {Purcell}, {Lumsden},
  {Oudmaijer}, {Moore}, {Busfield}, {Mottram}, \& {Davies}}]{urquhart09}
---. 2009, \aap, 501, 539

\bibitem[{{Urquhart} {et~al.}(2011){Urquhart}, {Morgan}, {Figura}, {Moore},
  {Lumsden}, {Hoare}, {Oudmaijer}, {Mottram}, {Davies}, \&
  {Dunham}}]{urquhart11}
---. 2011, \mnras, 418, 1689

\bibitem[{{Urquhart} {et~al.}(2012){Urquhart}, {Hoare}, {Lumsden}, {Oudmaijer},
  {Moore}, {Mottram}, {Cooper}, {Mottram}, \& {Rogers}}]{urquhart12}
---. 2012, \mnras, 420, 1656

\bibitem[{{Watson} {et~al.}(2003){Watson}, {Araya}, {Sewilo}, {Churchwell},
  {Hofner}, \& {Kurtz}}]{watson03}
{Watson}, C., {Araya}, E., {Sewilo}, M., {Churchwell}, E., {Hofner}, P., \&
  {Kurtz}, S. 2003, \apj, 587, 714

\bibitem[{{Watson} {et~al.}(2009){Watson}, {Corn}, {Churchwell}, {Babler},
  {Povich}, {Meade}, \& {Whitney}}]{watson09}
{Watson}, C., {Corn}, T., {Churchwell}, E.~B., {Babler}, B.~L., {Povich},
  M.~S., {Meade}, M.~R., \& {Whitney}, B.~A. 2009, \apj, 694, 546

\bibitem[{{Watson} {et~al.}(2008){Watson}, {Povich}, {Churchwell}, {Babler},
  {Chunev}, {Hoare}, {Indebetouw}, {Meade}, {Robitaille}, \&
  {Whitney}}]{watson08}
{Watson}, C., {et~al.} 2008, \apj, 681, 1341

\bibitem[{{Wienen} {et~al.}(2012){Wienen}, {Wyrowski}, {Schuller}, {Menten},
  {Walmsley}, {Bronfman}, \& {Motte}}]{wienen12}
{Wienen}, M., {Wyrowski}, F., {Schuller}, F., {Menten}, K.~M., {Walmsley},
  C.~M., {Bronfman}, L., \& {Motte}, F. 2012, \aap, 544, A146

\bibitem[{{Wilson} {et~al.}(1970){Wilson}, {Mezger}, {Gardner}, \&
  {Milne}}]{wilson70}
{Wilson}, T.~L., {Mezger}, P.~G., {Gardner}, F.~F., \& {Milne}, D.~K. 1970,
  \aap, 6, 364

\bibitem[{{Wink} {et~al.}(1982){Wink}, {Altenhoff}, \& {Mezger}}]{wink82}
{Wink}, J.~E., {Altenhoff}, W.~J., \& {Mezger}, P.~G. 1982, \aap, 108, 227

\bibitem[{{Wood} \& {Churchwell}(1989)}]{wc89a}
{Wood}, D.~O.~S., \& {Churchwell}, E. 1989, \apjs, 69, 831

\bibitem[{{Wouterloot} \& {Brand}(1989)}]{wouterloot89}
{Wouterloot}, J.~G.~A., \& {Brand}, J. 1989, \aaps, 80, 149

\bibitem[{{Wu} {et~al.}(2012){Wu}, {Xu}, {Menten}, {Zheng}, \& {Reid}}]{wu12}
{Wu}, Y.~W., {Xu}, Y., {Menten}, K.~M., {Zheng}, X.~W., \& {Reid}, M.~J. 2012,
  in IAU Symposium, Vol. 287, IAU Symposium, ed. R.~S. {Booth}, W.~H.~T.
  {Vlemmings}, \& E.~M.~L. {Humphreys}, 425--426

\bibitem[{{Xu} {et~al.}(2011){Xu}, {Moscadelli}, {Reid}, {Menten}, {Zhang},
  {Zheng}, \& {Brunthaler}}]{xu11}
{Xu}, Y., {Moscadelli}, L., {Reid}, M.~J., {Menten}, K.~M., {Zhang}, B.,
  {Zheng}, X.~W., \& {Brunthaler}, A. 2011, \apj, 733, 25

\bibitem[{{Xu} {et~al.}(2009){Xu}, {Reid}, {Menten}, {Brunthaler}, {Zheng}, \&
  {Moscadelli}}]{xu09}
{Xu}, Y., {Reid}, M.~J., {Menten}, K.~M., {Brunthaler}, A., {Zheng}, X.~W., \&
  {Moscadelli}, L. 2009, \apj, 693, 413

\bibitem[{{Xu} {et~al.}(2006){Xu}, {Reid}, {Zheng}, \& {Menten}}]{xu06}
{Xu}, Y., {Reid}, M.~J., {Zheng}, X.~W., \& {Menten}, K.~M. 2006, Science, 311,
  54

\bibitem[{{Xu} {et~al.}(2013){Xu}, {Li}, {Reid}, {Menten}, {Zheng},
  {Brunthaler}, {Moscadelli}, {Dame}, \& {Zhang}}]{xu13}
{Xu}, Y., {et~al.} 2013, \apj, 769, 15

\bibitem[{{Zhang} {et~al.}(2009){Zhang}, {Zheng}, {Reid}, {Menten}, {Xu},
  {Moscadelli}, \& {Brunthaler}}]{zhang09}
{Zhang}, B., {Zheng}, X.~W., {Reid}, M.~J., {Menten}, K.~M., {Xu}, Y.,
  {Moscadelli}, L., \& {Brunthaler}, A. 2009, \apj, 693, 419

\end{thebibliography}

\clearpage
%%%%%%%%%%%%%%%%%%%%%%%%%%%%%%%%%%%%%%%%%%%%%%%%%%
%%%%%%%%%%%%%%%%%%%%%%%%%%%%%%%%%%%%%%%%%%%%%%%%%%

\clearpage
%%%%%%%%%%%%%%%%%%%%%%%%%%%%%%%%%%%%%%%%%%%%%%%%%%
%%%%%%%%%%%%%%%%%%%%%%%%%%%%%%%%%%%%%%%%%%%%%%%%%%

\clearpage
%%%%%%%%%%%%%%%%%%%%%%%%%%%%%%%%%%%%%%%%%%%%%%%%%%
%%%%%%%%%%%%%%%%%%%%%%%%%%%%%%%%%%%%%%%%%%%%%%%%%%

%\clearpage
%%%%%%%%%%%%%%%%%%%%%%%%%%%%%%%%%%%%%%%%%%%%%%%%%%
%\begin{figure}[!ht]
%\begin{centering}
%\includegraphics[width=6.5in]{warp.eps}
%\caption{Galactic warp in the \hi\ layer as seen from the Sun.
%  Shown are the mean observed locations of the \hi\ layer for Galactic
%  radii of 10\,kpc, 12\,kpc, 16\,kpc, and 20\,kpc according to the
%  model in \citet{binney98}.  
%\hii\ regions observed in RRL emission
%  are over-plotted.  
%Observations of the outer disk of the Galaxy in the first Galactic quadrant
%  would require a significant extension in Galactic latitude from the
%  zone observed by MIPSGAL.}
%\label{fig:warp}
%\end{centering}
%\end{figure}
%%%%%%%%%%%%%%%%%%%%%%%%%%%%%%%%%%%%%%%%%%%%%%%%%%

\clearpage
%%%%%%%%%%%%%%%%%%%%%%%%%%%%%%%%%%%%%%%%%%%%%%%%%%
%%%%%%%%%%%%%%%%%%%%%%%%%%%%%%%%%%%%%%%%%%%%%%%%%%

\clearpage
%%%%%%%%%%%%%%%%%%%%%%%%%%%%%%%%%%%%%%%%%%%%%%%%%%
%%%%%%%%%%%%%%%%%%%%%%%%%%%%%%%%%%%%%%%%%%%%%%%%%%

\clearpage
%%%%%%%%%%%%%%%%%%%%%%%%%%%%%%%%%%%%%%%%%%%%%%%%%%
%%%%%%%%%%%%%%%%%%%%%%%%%%%%%%%%%%%%%%%%%%%%%%%%%%

\clearpage
%%%%%%%%%%%%%%%%%%%%%%%%%%%%%%%%%%%%%%%%%%%%%%%%%%
%%%%%%%%%%%%%%%%%%%%%%%%%%%%%%%%%%%%%%%%%%%%%%%%%%

\clearpage
%%%%%%%%%%%%%%%%%%%%%%%%%%%%%%%%%%%%%%%%%%%%%%%%%%
%%%%%%%%%%%%%%%%%%%%%%%%%%%%%%%%%%%%%%%%%%%%%%%%%%

\clearpage
%%%%%%%%%%%%%%%%%%%%%%%%%%%%%%%%%%%%%%%%%%%%%%%%%%
%%%%%%%%%%%%%%%%%%%%%%%%%%%%%%%%%%%%%%%%%%%%%%%%%%

\clearpage
%%%%%%%%%%%%%%%%%%%%%%%%%%%%%%%%%%%%%%%%%%%%%%%%%%
%%%%%%%%%%%%%%%%%%%%%%%%%%%%%%%%%%%%%%%%%%%%%%%%%%

\clearpage
%%%%%%%%%%%%%%%%%%%%%%%%%%%%%%%%%%%%%%%%%%%%%%%%%%
%%%%%%%%%%%%%%%%%%%%%%%%%%%%%%%%%%%%%%%%%%%%%%%%%%

%\clearpage
%%%%%%%%%%%%%%%%%%%%%%%%%%%%%%%%%%%%%%%%%%%%%%%%%%%
%\begin{figure}[!ht]
%\begin{centering}
%\includegraphics[width=6.5in]{G120_line.eps}
%\caption{Line of star formation????}
%\label{fig:G120}
%\end{centering}
%\end{figure}
%%%%%%%%%%%%%%%%%%%%%%%%%%%%%%%%%%%%%%%%%%%%%%%%%%%

\clearpage
%%%%%%%%%%%%%%%%%%%%%%%%%%%%%%%%%%%%%%%%%%%%%%%%%%
\begin{deluxetable}{lccccc}
\tabletypesize{\scriptsize}
\tablecaption{Radio Continuum Surveys}
\tablewidth{0pt}
\tablehead{
\colhead{Survey} &
\colhead{Wavelength} &
\colhead{Longitudes Used} &
\colhead{Latitudes Used} &
\colhead{Resolution} &
\colhead{Reference$^{\rm a}$}
\\
\colhead{} &
\colhead{cm} &
\colhead{} &
\colhead{} &
\colhead{arcsec} &
\colhead{}
}
\startdata
MAGPIS & \phn 6 & $350\degree < \ell < 42\degree\phn$ & $|b| \le 0\fdg4$                             & \phn 4  & 1 \\
MAGPIS & 20     & $\phn\phn 5\degree < \ell < 48\degree\phn$   & $|b| \le 0\fdg8$                    & \phn 6  & 2 \\
VGPS   & 21     & $\phn 18\degree < \ell < 66\degree\phn$      & $|b|$ from $1\degree$ to $2\degree$ & \phn 60 & 3 \\
CGPS   & 21     & $\phn 66\degree < \ell < 175\degree$         & $-3.5\degree < b < +5.5\degree$    & \phn 60 & 4 \\
NVSS   & 20     & $\phn 66\degree < \ell < 247\degree$         & $|b| \le 8\degree$                    & \phn 45 & 5 \\
SGPS   & 21     & $253\degree < \ell < 358\degree$             & $|b| \le 1\fdg5$                       & 120     & 6 \\
SUMSS  & 36     & $247\degree < \ell < 337\degree$             & $|b| \le 8\degree$                  & \phn 45 & \,\,7
\tablenotetext{a}{1:~\citet{becker94}; 2:~\citet{helfand06}; 3:
  \citet{stil06}; 4:~\citet{taylor03}; 5:~\citet{condon98}; 6:
  \citet{mcclure05}; 7:~\citet{bock99}}

\enddata
\label{tab:continuum}
\end{deluxetable}
%%%%%%%%%%%%%%%%%%%%%%%%%%%%%%%%%%%%%%%%%%%%%%%%%%

\clearpage
%%%%%%%%%%%%%%%%%%%%%%%%%%%%%%%%%%%%%%%%%%%%%%%%%%
\begin{deluxetable}{lcrrrllrrccccccccc}
\renewcommand{\tabcolsep}{3pt}
\tabletypesize{\tiny}
\tablecaption{WISE Catalog of Galactic \hii\ Regions}
\tablewidth{0pt}
\tablehead{
\colhead{WISE Name} &
\colhead{Cat.} &
\colhead{$\ell$} &
\colhead{$b$} &
\colhead{Radius} &
\colhead{HII Region Name} &
\colhead{Membership} &
\colhead{$\ell$} &
\colhead{$b$} &
\colhead{V$_{\rm LSR}$} &
\colhead{$\sigma_{\rm VLSR}$} &
\colhead{$\Delta V$} &
\colhead{$\sigma_{\Delta V}$} &
\colhead{Ref.$^{\rm a}$}
\\
\colhead{} &
\colhead{} &
\colhead{deg.} &
\colhead{deg.} &
\colhead{arcsec.} &
\colhead{} &
\colhead{} &
\colhead{deg.} &
\colhead{deg.} &
\colhead{\kms} &
\colhead{\kms} &
\colhead{\kms} &
\colhead{\kms} &
\colhead{} &
}
\startdata
G029.950$-$00.809& K&  29.951&  $-$0.809&   64& G029.951$-$00.800& G029.951$-$00.800&  29.951&  $-$0.800&                  86.1&             0.4&                21.3&             0.9&  2\\
G029.954$-$00.788& G&  29.954&  $-$0.787&   29&                & G029.951$-$00.800&        &        &                      &                &                    &                &   \\
G029.955$-$00.093& G&  29.956&  $-$0.092&  124&                & G029.960$-$00.020&        &        &                      &                &                    &                &   \\
G029.956$-$00.020& K&  29.956&  $-$0.019&   94& G029.960$-$00.020& G029.960$-$00.020&  29.956&  $-$0.017&                  91.3&             0.1&                32.2&             0.3& 10\\
G029.959$-$00.348& Q&  29.959&  $-$0.348&   37&                &                &        &        &                      &                &                    &                &   \\
G029.970$-$00.412& Q&  29.970&  $-$0.411&   14&                &                &        &        &                      &                &                    &                &   \\
G029.970$-$00.417& Q&  29.970&  $-$0.416&   14&                &                &        &        &                      &                &                    &                &   \\
G029.975$-$00.835& G&  29.975&  $-$0.835&   72&                & G029.951$-$00.800&        &        &                      &                &                    &                &   \\
G029.976+00.547& Q&  29.976&   0.548&  137&                &                &        &        &                      &                &                    &                &   \\
G029.978+00.020& C&  29.979&   0.021&   30&                &                &        &        &                      &                &                    &                &   \\
G029.979$-$00.587& K&  29.980&  $-$0.587&  219& G030.001$-$00.602&                &  30.001&  $-$0.602&             42.6;93.2&         1.8;1.1&           11.2;21.9&         4.3;2.5&  9\\
G029.997+00.072& Q&  29.997&   0.073&   39&                &                &        &        &                      &                &                    &                &   \\
G030.003$-$00.267& C&  30.004&  $-$0.266&   24&                &                &        &        &                      &                &                    &                &   \\
G030.007$-$00.084& G&  30.008&  $-$0.084&   82&                & G029.960$-$00.020&        &        &                      &                &                    &                &   \\
G030.009$-$00.274& Q&  30.010&  $-$0.273&   24&                &                &        &        &                      &                &                    &                &   \\
G030.014+00.017& Q&  30.014&   0.017&   23&                &                &        &        &                      &                &                    &                &   \\
G030.015+00.025& Q&  30.015&   0.025&   22&                &                &        &        &                      &                &                    &                &   \\
G030.022$-$00.042& K&  30.022&  $-$0.042&   60& G030.022$-$00.044&                &  30.022&  $-$0.044&                  94.9&             0.1&                21.7&             0.1&  2\\
G030.026+00.109& K&  30.026&   0.109&   52& G030.024+00.110&                &  30.024&   0.110&                 100.6&             0.1&                26.5&             0.3&  2\\
G030.030+00.005& Q&  30.031&   0.005&   38&                &                &        &        &                      &                &                    &                &   \\
G030.036$-$00.167& K&  30.037&  $-$0.166&  265& G030.069$-$00.160&                &  30.069&  $-$0.160&                  98.2&             1.7&                32.1&             4.1&  8\\
G030.042$-$00.144& C&  30.043&  $-$0.143&   48&                &                &        &        &                      &                &                    &                &   \\
G030.051+00.180& C&  30.051&   0.180&  150&                &                &        &        &                      &                &                    &                &   \\
G030.055$-$00.339& C&  30.055&  $-$0.339&   68&                &                &        &        &                      &                &                    &                &   \\
G030.057+00.077& Q&  30.057&   0.078&   39&                &                &        &        &                      &                &                    &                &   \\

\enddata
\label{tab:catalog}
\tablenotetext{a}{1:~\citet{arvidsson09}; 2:~\citet{anderson11}; 3:~\citet{araya02};
4:~\citet{bania12}; 5:~\citet{caswell87}; 6:~\citet{downes80}; 7:~\citet{fich90}; 8:~\citet{lockman89};
9:~\citet{lockman96}; 10:~\citet{sewilo04b}; 11:~\citet{watson03}; 12:~\citet{wilson70}; 13:~\citet{wink82}}
\tablenotetext{\,\!}{Note: If there are multiple WISE sources within the beam of an observed position, all such WISE sources
share the same line parameters.}
\tablenotetext{\,\!}{Only a portion of the table is shown here to
  demonstrate its form and content.  A machine-readable version of the
  full table is available.}

\end{deluxetable}
%%%%%%%%%%%%%%%%%%%%%%%%%%%%%%%%%%%%%%%%%%%%%%%%%%

\clearpage
%%%%%%%%%%%%%%%%%%%%%%%%%%%%%%%%%%%%%%%%%%%%%%%%%%
\begin{deluxetable}{lrrcl}
\tabletypesize{\scriptsize}
\tablecaption{Sources not Included in the Catalog}
\tablewidth{0pt}
\tablehead{
\colhead{Name} &
\colhead{$\ell$} &
\colhead{$b$} &
\colhead{Author$^{\rm a}$} &
\colhead{Reason$^{\rm b}$}
\\
\colhead{} &
\colhead{deg.} &
\colhead{deg.} &
\colhead{} &
\colhead{}
}
\startdata
G001.435$-$00.131	&	1.435	&	$-$0.131	&	6	&	No IR			\\
G002.370$+$01.400	&	2.370	&	1.400		&	3	&	SNR			\\
G003.262$+$00.019	&	3.262	&	0.019		&	5	&	Not distinct	  	\\
G006.565$-$00.297	&	6.565	&	$-$0.297	&	5	&	No IR			\\
G006.616$-$00.545	&	6.616	&	$-$0.545	&	6	&	No IR			\\
G006.667$-$00.247	&	6.667	&	$-$0.247	&	5	&	No IR			\\
G007.002$-$00.015	&	7.002	&	$-$0.015	&	6	&	Not distinct		\\
G007.236$+$00.144	&	7.236	&	0.144		&	6	&	Not distinct		\\
G008.094$+$00.085	&	8.094	&	0.085		&	6	&	Not distinct		\\
G008.415$+$00.033	&	8.415	&	0.033		&	6	&	Not distinct		\\
G008.786$-$00.034	&	8.786	&	$-$0.034	&	6	&	No IR			\\
G009.741$+$00.842	&	9.741	&	0.842		&	1	&	Probable PN		\\
G011.207$+$00.088	&	11.207	&	0.088		&	5	&	No IR	 		\\
G012.114$-$00.447	&	12.114	&	$-$0.447	&	6	&	No IR			\\
G012.116$+$00.076	&	12.116	&	0.076		&	1	&	Probable PN		\\
G012.325$-$00.305	&	12.325	&	$-$0.305	&	6	&	No IR	 		\\
G012.575$+$00.216	&	12.575	&	0.216		&	6	&	No IR			\\
G014.081$-$00.151	&	14.081	&	$-$0.151	&	5	&	SNR$^{\rm c}$		\\
G014.161$-$00.223	&	14.161	&	$-$0.223	&	5	&	SNR$^{\rm c}$		\\
G014.221$-$00.287	&	14.221	&	$-$0.287	&	5	&	SNR$^{\rm c}$		\\
G014.330$-$00.183	&	14.330	&	$-$0.183	&	5	&	SNR$^{\rm c}$		\\
G014.380$-$00.549	&	14.380	&	$-$0.549	&	6	&	No IR			\\
G014.600$+$00.014	&	14.600	&	0.014		&	5	&	Not distinct		\\
G014.931$-$00.638	&	14.931	&	$-$0.638	&	5	&	Not distinct		\\
G015.125$-$00.529	&	15.125	&	$-$0.529	&	1	&	Not distinct		\\
G015.143$-$00.940	&	15.143	&	$-$0.940	&	5	&	Not distinct		\\
G015.181$-$00.625	&	15.181	&	$-$0.625	&	5	&	Not distinct		\\
G015.198$-$00.768	&	15.198	&	$-$0.768	&	5	&	Not distinct		\\
G015.452$+$00.191	&	15.452	&	0.191		&	6	&	SNR$^{\rm c}$		\\
G015.520$-$00.143	&	15.520	&	$-$0.143	&	6	&	SNR$^{\rm c}$		\\
G016.228$-$00.368	&	16.228	&	$-$0.368	&	1	&	Not distinct		\\
G016.404$-$00.575	&	16.404	&	$-$0.575	&	1	&	Probable PN		\\
G016.463$+$00.966	&	16.463	&	0.966		&	6	&	No IR	 		\\
G016.984$+$00.934	&	16.984	&	0.934		&	5	&	Not distinct		\\
G017.144$+$00.765	&	17.144	&	0.765		&	5	&	Not distinct		\\
G017.226$+$00.390	&	17.226	&	0.390		&	6	&	Evolved star StRS 177	\\
G017.250$-$00.195	&	17.250	&	$-$0.195	&	6	&	No IR			\\
G017.364$+$00.519	&	17.364	&	0.519		&	1	&	Probable PN		\\
G018.643$-$00.288	&	18.643	&	$-$0.288	&	5	&	No IR	 		\\
G018.708$-$00.126	&	18.708	&	$-$0.126	&	1	&	Probable PN		\\
G020.264$-$00.894	&	20.264	&	$-$0.894	&	5	&	No IR	 		\\
G020.479$+$00.165	&	20.479	&	0.165		&	5	&	SNR$^{\rm c, f}$	\\
G020.681$-$00.136	&	20.681	&	$-$0.136	&	5	&	Not distinct		\\
G021.666$-$00.261	&	21.666	&	$-$0.261	&	6	&	No IR			\\
G022.039$+$00.045	&	22.039	&	0.045		&	6	&	No IR			\\
G022.154$-$00.153	&	22.154	&	$-$0.153	&	1	&	Probable PN		\\
G022.162$-$00.157	&	22.162	&	$-$0.157	&	6	&	High-Mass X-ray Binary Candidate 	\\
G022.739$-$00.303	&	22.739	&	$-$0.303	&	1	&	Not distinct	    	\\
G022.935$-$00.072	&	22.935	&	$-$0.072	&	6	&	No IR			\\
G023.000$+$00.219	&	23.000	&	0.219		&	6	&	No IR			\\
G023.067$-$00.367	&	23.067	&	$-$0.367	&	5	&	No IR			\\
G023.072$-$00.248	&	23.072	&	$-$0.248	&	5	&	No IR			\\
G023.162$+$00.023	&	23.162	&	0.023		&	6	&	No IR			\\
G023.254$-$00.268	&	23.254	&	$-$0.268	&	5	&	Not distinct		\\
G023.849$-$00.196	&	23.849	&	$-$0.196	&	1	&	Not distinct		\\
G024.430$-$00.320	&	24.430	&	$-$0.320	&	1	&	Not distinct		\\
G024.775$-$00.400	&	24.775	&	$-$0.400	&	6	&	No IR			\\
G025.405$-$00.254	&	25.405	&	$-$0.254	&	5	&	Not distinct		\\
G025.945$+$00.125	&	25.945	&	0.125		&	6	&	No IR			\\
G026.467$+$00.018	&	26.467	&	0.018		&	6	&	WR or LBV star 2MASS J18393224-0544204$^{\rm d}$	\\
G026.510$+$00.280	&	26.508	&	0.283		&	7	&	No Radio		\\
G026.863$-$00.275	&	26.863	&	$-$0.275	&	1	&	No IR			\\
G027.132$+$00.000	&	27.132	&	0.000		&	5	&	SNR$^{\rm f}$		\\
G029.433$+$00.273	&	29.433	&	0.273		&	1	&	No IR			\\
G029.551$+$00.114	&	29.551	&	0.114		&	6	&	SNR$^{\rm e}$		\\
G029.874$-$00.819	&	29.874	&	$-$0.819	&	1	&	Probable PN		\\
G030.277$-$00.020	&	30.277	&	$-$0.020	&	5	&	No IR	 		\\
G030.509$-$00.447	&	30.509	&	$-$0.447	&	5	&	No IR			\\
G030.532$-$00.259	&	30.532	&	$-$0.259	&	1	&	Probable PN		\\
G030.663$-$00.334	&	30.663	&	$-$0.334	&	1	&	Probable PN		\\
G030.771$-$00.223	&	30.771	&	$-$0.223	&	1	&	Not distinct		\\
G030.820$+$00.270	&	30.818	&	0.274		&	7	&	No radio		\\
G030.832$-$00.186	&	30.832	&	$-$0.186	&	5	&	Not distinct		\\
G030.840$-$00.110	&	30.839	&	$-$0.113	&	7	&	Not distinct		\\
G030.854$+$00.134	&	30.854	&	0.134		&	5	&	SNR$^{\rm f}$		\\
G031.607$+$00.334	&	31.607	&	0.334		&	6	&	SNR$^{\rm e}$		\\
G031.650$-$00.649	&	31.650	&	$-$0.649	&	5	&	No IR			\\
G031.671$+$00.238	&	31.671	&	0.238		&	1	&	Not distinct		\\
G035.588$-$00.489	&	35.588	&	$-$0.489	&	5	&	Probable PN		\\
G037.439$-$00.040	&	37.439	&	$-$0.040	&	5	&	Not distinct		\\
G038.305$+$00.139	&	38.305	&	0.139		&	6	&	No IR			\\
G046.017$+$00.264	&	46.017	&	0.264		&	1	&	Probable PN		\\
G046.017$+$00.264	&	46.017	&	0.264		&	2	&	Probable PN		\\
G050.232$+$00.326	&	50.232	&	0.326		&	5	&	No IR	 		\\
G050.556$+$00.045	&	50.556	&	0.045		&	1	&	Probable PN		\\
G051.362$-$00.001	&	51.362	&	$-$0.001	&	5	&	Not distinct		\\
G075.363$-$00.423	&	75.363	&	$-$0.423	&	5	&	Not distinct		\\
G078.303$+$02.463	&	78.303	&	2.463		&	6	&	Not distinct		\\
G078.341$+$01.765	&	78.341	&	1.765		&	6	&	Not distinct		\\
G078.374$+$02.520	&	78.374	&	2.520		&	6	&	Not distinct		\\
G078.455$-$00.043	&	78.455	&	$-$0.043	&	5	&	Not distinct		\\
G078.637$+$00.847	&	78.637	&	0.847		&	5	&	Not distinct		\\
G078.653$+$02.041	&	78.653	&	2.041		&	6	&	Not distinct		\\
G078.683$+$00.939	&	78.683	&	0.939		&	5	&	Not distinct		\\
G078.727$-$00.606	&	78.727	&	$-$0.606	&	5	&	Not distinct		\\
G079.235$+$00.342	&	79.235	&	0.342		&	5	&	Not distinct		\\
G079.418$+$02.417	&	79.418	&	2.417		&	5	&	Not distinct		\\
G079.957$+$00.866	&	79.957	&	0.866		&	5	&	Not distinct		\\
G079.982$-$00.028	&	79.982	&	$-$0.028	&	5	&	Not distinct		\\
G081.551$+$02.074	&	81.551	&	2.074		&	5	&	Not distinct		\\
G082.277$+$02.425	&	82.277	&	2.425		&	5	&	Not distinct		\\
G082.454$+$02.369	&	82.454	&	2.369		&	5	&	Not distinct		\\
G085.178$-$00.074	&	85.178	&	$-$0.074	&	5	&	Not distinct		\\
BFS~53			&	211.290	&	$-$1.085	&	4	&	No IR			\\
BFS~55			&	216.088	&	0.017		&	4	&	Not distinct		\\
BFS~56			&	217.298	&	$-$0.039	&	4	&	No Radio		\\
G283.524$-$01.004	&	283.524	&	$-$1.004	&	3	&	Not distinct		\\
G284.260$+$00.400	&	284.260	&	0.400		&	3	&	No IR			\\
G284.559$-$00.183	&	284.559	&	$-$0.183	&	3	&	Not distinct		\\
G286.873$-$00.883	&	286.873	&	$-$0.883	&	3	&	No IR			\\
G287.217$+$00.053	&	287.217	&	0.053		&	3	&	No IR			\\
G287.550$-$00.616	&	287.550	&	$-$0.616	&	3	&	Not distinct		\\
G287.636$-$00.936	&	287.636	&	$-$0.936	&	3	&	Not distinct		\\
G287.956$-$00.823	&	287.956	&	$-$0.823	&	3	&	Not distinct		\\
G290.487$-$00.814	&	290.487	&	$-$0.814	&	3	&	No IR			\\
G305.097$+$00.138	&	305.097	&	0.138		&	3	&	Not distinct		\\
G306.256$+$00.066	&	306.256	&	0.066		&	3	&	Not distinct		\\
G311.288$-$00.021	&	311.288	&	$-$0.021	&	3	&	Not distinct		\\
G317.037$+$00.300	&	317.037	&	0.300		&	3	&	Not distinct		\\
G326.141$-$00.328	&	326.141	&	$-$0.328	&	3	&	Not distinct		\\
G326.230$+$00.976	&	326.230	&	0.976		&	3	&	Not distinct		\\
G336.840$+$00.047	&	336.840	&	0.047		&	3	&	Not distinct		\\
G345.031$+$01.540	&	345.031	&	1.540		&	3	&	Not distinct		\\
G345.283$-$00.990	&	345.283	&	$-$0.990	&	1	&	Not distinct		\\
G347.386$+$00.266	&	347.386	&	0.266		&	3	&	Not distinct		\\
G347.536$+$00.084	&	347.536	&	0.084		&	1	&	Not distinct		\\
G348.607$-$00.551	&	348.607	&	$-$0.551	&	3	&	Not distinct		\\
G349.804$-$00.176	&	349.804	&	$-$0.176	&	1	&	No IR			\\
G352.676$+$00.148	&	352.676	&	0.148		&	3	&	PN H 1$-$12		\\
G355.581$+$00.288	&	355.581	&	0.288		&	1	&	Not distinct		\\
G358.623$-$00.066	&	358.623	&	$-$0.066	&	3	&	Not distinct		\\
G359.158$-$00.881	&	359.158	&	$-$0.881	&	6	&	No IR			\\
G359.186$-$00.026	&	359.186	&	$-$0.026	&	3	&	Not distinct		\\

\tablenotetext{a}{1:~\citet{anderson11}; 2:~\citet{bania12};
  3:~\citet{caswell87}; 4:~\citet{fich90}; 5:~\citet{lockman89};
  6:~\citet{lockman96}; 7:~\citet{sewilo04b}}

\tablenotetext{b}{``No IR'': Infrared emission absent or morphology
  not characteristic of HII regions; ``Not distinct'': The location
  observed is not distinct from a known HII region that is included in
  the catalog; ``Probable PN'': The object has compact MIR emission
  with no nebulosity; ``No radio'': No discrete radio continuum source
  at the observed position}

\tablenotetext{c}{\citet{brogan06}}
\tablenotetext{d}{\citet{clark03}}
\tablenotetext{e}{\citet{gaensler99}}
\tablenotetext{f}{\citet{helfand06}}
%\tablenotetext{g}{\citet{kwok97}}
%\tablenotetext{h}{\citet{stephenson92}}
\enddata
\label{tab:removed}
\end{deluxetable}
%%%%%%%%%%%%%%%%%%%%%%%%%%%%%%%%%%%%%%%%%%%%%%%%%%

\clearpage
%%%%%%%%%%%%%%%%%%%%%%%%%%%%%%%%%%%%%%%%%%%%%%%%%%
\begin{deluxetable}{lcrrrrrrrccc}
\tabletypesize{\tiny}
\renewcommand{\tabcolsep}{3pt}
\tablecaption{WISE Correlations with Dust Continuum and Molecular Line Surveys}
\tablewidth{0pt}
\tablehead{
\colhead{Survey} &
\colhead{Approx. Longitude Range} &
\colhead{K} &
\colhead{G} &
\colhead{C} &
\colhead{Q} &
\colhead{Total$^{\rm a}$} &
\colhead{\% WISE$^{\rm b}$} &
\colhead{\% Survey} &
\colhead{Wavelength} &
\colhead{Molecule} &
\colhead{Reference$^{\rm c}$}
}
\startdata
ATLASGAL &	          295 &	           84 &	          283 &	          728 &	         1395 &	           49 &	           24 &	 870\,\micron              &	 \nodata &	 1 \\
BGPS 	 &	          501 &	          161 &	          295 &	          809 &		 1766 &	           54 &	           21 &	 1.1\,mm 		   &	 \nodata &	 2 \\
MWP (Large) &	          708 &	          158 &	          554 &	          581 &		 2001 &	           33 &	  	   53 &	 8\,\micron\ + 24\,\micron &	 \nodata &	 3 \\
MWP (Small) &	           82 &	           32 &	          117 &	          433 &		  664 &	           11 &	  	   49 &	 8\,\micron\ + 24\,\micron &	 \nodata &	 3 \\
RMS (HII Regions) &       225 &	           95 &	          153 &	          127 &		  608 &	           18 &	           75 &	 8--21\,\micron 	   &	 \nodata &	 4 \\
RMS (YSOs) &	           39 &	           26 &	           50 &	          176 &		  299 &	            9 &	           44 &	 8--21\,\micron 	   &	 \nodata &	 4 \\
All Continuum &	   	 1049 &	          361 &	          982 &	         1941 &	         4349 &	           60 &	      \nodata &       		   \nodata &	 \nodata & \nodata \\ \hline

ATLASGAL &	          109 &	           48 &	           43 &	           87 &	          287 &	           13 &	           59 &	 \nodata       &	 \cor    &	 5 \\
ATLASGAL &	           40 &	           11 &	           22 &	           60 &		  133 &	            5 &	           18 &	 \nodata       &	 NH$_3$  &	 5 \\
BGPS 	 &	          239 &	           84 &	          122 &	          209 &		  654 &	           18 &	           47 &	 \nodata       &	 HCO$^+$ &	 6 \\
BGPS 	 &	           35 &	            9 &	           18 &	           29 &	           91 &	            2 &	            7 &	 \nodata       &	 NH$_3$  &	 7 \\
GRS  		&	  237 &	            0 &	            0 &	            0 &		  237 &	           11 &	          105 &	 \nodata       &	 \cor 	 &	 8 \\
HOPS &	           	   35 &	            7 &	           33 &	           52 &	 	  127 &	            3 &	           18 &	 \nodata       &	 NH$_3$  &	 9 \\ 
IRAS &	           	   28 &	           21 &	          115 &	          253 &	 	  434 &	           17 &	           35 &	 \nodata       &	 \cor 	 &	10 \\
IRAS &	          	  307 &	           69 &	          226 &	          202 &	 	  812 &	           11 &	           56 &	 \nodata       &	 CS 	 &	11 \\
Malt90 Pilot &	           12 &	            9 &	           35 &	           31 &		   88 &	            3 &	           48 &	 \nodata       &	 Many 	 &	12 \\
RMS &	          	  165 &	           87 &	           58 &	          104 &		  414 &	           12 &	           45 &	 \nodata       &	 \cor 	 &	13 \\
RMS &	           	   53 &	           34 &	          100 &	          114 &		  315 &	           12 &	           37 &	 \nodata       &	 \cor 	 &	14 \\
RMS &	          	  145 &	           71 &	           63 &	          121 &		  400 &	           11 &	           68 &	 \nodata       &	 NH$_3$  &	15 \\
All Molecular &	          580 &	          546 &	          546 &	          851 &		 2243 &	           31 &	      \nodata &	 \nodata       &	 \nodata &	 \nodata \\

\enddata
\label{tab:correlations}
\tablenotetext{a}{Also includes sources with no radio data.}

\tablenotetext{b}{Only computed over respective survey areas, except
  for the ``All Continuum'' and ``All Molecular'' columns.}

\tablenotetext{c}{1:~\citet{contreras13};
  2:~\citet{rosolowsky10}; 3:~\citet{simpson12};
  4:~\citet{urquhart08a}; catalog retrieved from
    http://www.ast.leeds.ac.uk/cgi-bin/RMS/RMS\_DATABASE.cgi;
  5:~\citet{wienen12}; 6:~\citet{schlingman11}; 7:~\citet{dunham11};
  8:~\citet{anderson09b}; 9:~\citet{purcell13}; 10:~\citet{bronfman96};
  11:~\citet{wouterloot89}; 12:~\citet{foster11};
  13:~\citet{urquhart08b}; 14:~\citet{urquhart07b};
  15:~\citet{urquhart11}}

%\tablenotetext{d}{Only sources identified as YSOs or HII regions.}

\end{deluxetable}
%%%%%%%%%%%%%%%%%%%%%%%%%%%%%%%%%%%%%%%%%%%%%%%%%%

\clearpage
%%%%%%%%%%%%%%%%%%%%%%%%%%%%%%%%%%%%%%%%%%%%%%%%%%
\begin{deluxetable}{llrrrcr}
\tabletypesize{\tiny}
\tablecaption{Molecular Line Velocities for WISE Sources}
\tablewidth{0pt}
\tablehead{
\colhead{WISE Name} &
\colhead{Source Name} &
\colhead{$\ell$} &
\colhead{$b$} &
\colhead{V$_{\rm LSR}$} &
\colhead{Molecule} &
\colhead{Ref.$^{\rm a}$}
\\
\colhead{} &
\colhead{} &
\colhead{deg.} &
\colhead{deg.} &
\colhead{\kms} &
\colhead{} &
\colhead{}
}
\startdata
G029.950$-$00.809&                      &        &        &                                &       &   \\
G029.954$-$00.788&       G029.959$-$00.791&  29.959&  $-$0.791&                            85.2&   HCO+&  6\\
G029.955$-$00.093&                      &        &        &                                &       &   \\
G029.956$-$00.020&     G029.9564$-$00.0174&  29.957&  $-$0.017&                            97.2&    NH3&  9\\
G029.959$-$00.348&                      &        &        &                                &       &   \\
G029.970$-$00.412&                      &        &        &                                &       &   \\
G029.970$-$00.417&                      &        &        &                                &       &   \\
G029.975$-$00.835&                      &        &        &                                &       &   \\
G029.976+00.547&                      &        &        &                                &       &   \\
G029.978+00.020&                      &        &        &                                &       &   \\
G029.979$-$00.587&           G30.00$-$0.60&  30.000&  $-$0.600&                            92.1&   13CO&  1\\
G029.997+00.072&                      &        &        &                                &       &   \\
G030.003$-$00.267&       G030.006$-$00.272&  30.006&  $-$0.272&                           101.8&   HCO+&  6\\
G030.007$-$00.084&                      &        &        &                                &       &   \\
G030.009$-$00.274&       G030.006$-$00.272&  30.006&  $-$0.272&                           101.8&   HCO+&  6\\
G030.014+00.017&       G030.022+00.009&  30.022&   0.009&                           105.1&   HCO+&  6\\
G030.015+00.025&                      &        &        &                                &       &   \\
G030.022$-$00.042&       G030.020$-$00.050&  30.020&  $-$0.050&                            91.7&   HCO+&  6\\
G030.026+00.109&       G030.026+00.106&  30.026&   0.106&                           106.0&   HCO+&  6\\
G030.030+00.005&       G030.022+00.009&  30.022&   0.009&                           105.1&   HCO+&  6\\
G030.036$-$00.167&                      &        &        &                                &       &   \\
G030.042$-$00.144&                      &        &        &                                &       &   \\
G030.051+00.180&                      &        &        &                                &       &   \\
G030.055$-$00.339&                      &        &        &                                &       &   \\
G030.057+00.077&                      &        &        &                                &       &   \\

\enddata
\label{tab:mol_lines}

\tablenotetext{a}{1:~\citet{anderson09b}; 2:~\citet{bronfman96};
  3:~\citet{dunham11}; 4:~\citet{foster11}; 5:~\citet{purcell12};
  6:~\citet{schlingman11}; 7:~\citet{urquhart07b};
  8:~\citet{urquhart08b}; 9:~\citet{wienen12};
  10:~\citet{wienen12}, \citet{jackson06}; 11:\citet{wouterloot89}}
\tablenotetext{\,}{Only a portion of the table is shown here to
  demonstrate its form and content.  A machine-readable version of the
  full table is available.}
\end{deluxetable}
%%%%%%%%%%%%%%%%%%%%%%%%%%%%%%%%%%%%%%%%%%%%%%%%%%

\clearpage
%%%%%%%%%%%%%%%%%%%%%%%%%%%%%%%%%%%%%%%%%%%%%%%%%%
\begin{deluxetable}{lcccccccrrrrc}
\tabletypesize{\tiny}
\tablecaption{Distances}
\tablewidth{0pt}
\tablehead{
\colhead{WISE Name} &
\colhead{Near} &
\colhead{Far} &
\colhead{Tangent} &
\colhead{R$_{\rm Gal}$} &
\colhead{V$_{\rm T}$} &
\colhead{KDAR} &
\colhead{d$_\sun$} &
\colhead{$\sigma_{\rm d\sun}$} &
\colhead{Azimuth} &
\colhead{z} &
\colhead{Method$^{\rm a}$} &
\colhead{Ref.$^{\rm b}$}
\\
\colhead{} &
\colhead{kpc} &
\colhead{kpc} &
\colhead{kpc} &
\colhead{kpc} &
\colhead{\kms} &
\colhead{} &
\colhead{kpc} &
\colhead{kpc} &
\colhead{deg.} &
\colhead{pc} &
\colhead{} &
\colhead{}
}
\startdata
G029.950$-$00.809&  5.3&  9.4& 7.4&  4.7&  107.5&  &     &     &    &      &                           &   \\
G029.954$-$00.788&  5.2&  9.5& 7.4&  4.8&  107.5&  &     &     &    &      &                           &   \\
G029.955$-$00.093&     &     &    &     &       &  &     &     &    &      &                           &   \\
G029.956$-$00.020&  5.6&  9.1& 7.4&  4.6&  107.5& F&  9.1&  0.6&  82&    $-$3&                   H2CO/Kin& 32\\
G029.959$-$00.348&     &     &    &     &       &  &     &     &    &      &                           &   \\
G029.970$-$00.412&     &     &    &     &       &  &     &     &    &      &                           &   \\
G029.970$-$00.417&     &     &    &     &       &  &     &     &    &      &                           &   \\
G029.975$-$00.835&     &     &    &     &       &  &     &     &    &      &                           &   \\
G029.976+00.547&     &     &    &     &       &  &     &     &    &      &                           &   \\
G029.978+00.020&     &     &    &     &       &  &     &     &    &      &                           &   \\
G029.979$-$00.587&  5.6&  9.1& 7.4&  4.6&  107.4& F&  9.1&  0.6&  82&   $-$93&              HIEA+HISA/Kin&  1\\
G029.997+00.072&     &     &    &     &       &  &     &     &    &      &                           &   \\
G030.003$-$00.267&  6.4&  8.4& 7.4&  4.4&  107.4& T&  7.4&  1.3&  59&   $-$34&               TP (Mol)/Kin&   \\
G030.007$-$00.084&     &     &    &     &       &  &     &     &    &      &                           &   \\
G030.009$-$00.274&  6.4&  8.4& 7.4&  4.4&  107.3& T&  7.4&  1.3&  59&   $-$35&               TP (Mol)/Kin&   \\
G030.014+00.017&  6.7&  8.0& 7.4&  4.3&  107.3& T&  7.4&  1.1&  59&     2&               TP (Mol)/Kin&   \\
G030.015+00.025&     &     &    &     &       &  &     &     &    &      &                           &   \\
G030.022$-$00.042&  5.8&  8.9& 7.4&  4.5&  107.3& F&  8.9&  0.6&  79&    $-$6&                   HIEA/Kin&  2\\
G030.026+00.109&  6.3&  8.5& 7.4&  4.4&  107.3& T&  7.4&  1.4&  59&    14&                   HIEA/Kin&  2\\
G030.030+00.005&  6.7&  8.0& 7.4&  4.3&  107.3& T&  7.4&  1.1&  59&     0&               TP (Mol)/Kin&   \\
G030.036$-$00.167&  6.1&  8.7& 7.4&  4.4&  107.3& F&  8.7&  0.8&  76&   $-$25&              HIEA+HISA/Kin&  1\\
G030.042$-$00.144&     &     &    &     &       &  &     &     &    &      &                           &   \\
G030.051+00.180&     &     &    &     &       &  &     &     &    &      &                           &   \\
G030.055$-$00.339&     &     &    &     &       &  &     &     &    &      &                           &   \\
G030.057+00.077&     &     &    &     &       &  &     &     &    &      &                           &   \\

\enddata
\label{tab:distances}

\tablenotetext{a}{``Parallax'' for maser parallax distances,
  ``OG/Kin.''  for outer Galaxy kinematic distances, ``TP/Kin'' for
  tangent point kinematic distances, ``HIEA+HISA/Kin'' for HIE/A and
  HISA kinematic distances from \citet{anderson09b}, ``HIEA/Kin'' for
  HIE/A kinematic distances, ``H$_2$CO/Kin.'' for H$_2$CO absorption
  kinematic distances, ``IRDC+HISA (Mol)/Kin'' for kinematic distances
  from \citet{dunham11},  ``Visible/Kin'' for inner Galaxy Sharpless
  HII region kinematic distances (near distances assumed). Kinematic
  distances derived from molecular line observations use the above
  designations, but include ``(Mol)'', i.e., ``OG~(Mol)/Kin''.}

\tablenotetext{b}{1:~\citet{anderson09b}; 2:~\citet{anderson12c};
  3:~\citet{ando11}; 4:~\citet{araya02}; 5:~\citet{bania12};
  6:~\citet{bartkiewicz08}; 7:~\citet{brunthaler09};
  8:~\citet{caswell87}; 9:~\citet{dunham11}; 10:~\citet{hachisuka09};
  11:~Hachisuka et al. (2013, in prep.); 12:~\citet{hirota08a};
  13:~\citet{hirota08b}; 14:~\citet{honma07}; 15:~\citet{immer13};
  16:~\citet{jones12}; 17:~Jones et al. (2013, in press); 18:~\citet{kurayama11};
  19:~\citet{menten07}; 20: \citet{moellenbrock09};
  21:~\citet{moscadelli09}; 22:~\citet{niinuma11}; 23:~\citet{oh10};
  24:~\citet{reid09a}; 25:~\citet{rygl10}; 26:~\citet{rygl12};
  27:~\citet{sanna09}; 28:~\citet{sanna12}; 29:~\citet{sato08};
  30:~\citet{sato10a}; 31:~\citet{sato10b}; 32:~\citet{sewilo04b};
  33:~\citet{shiozaki11}; 34:~\citet{urquhart12};
  35:~\citet{urquhart12}/\citet{roman-duval09}; 36:~\citet{watson03};
  37:~\citet{wu12}; 38:~\citet{xu06}; 39:~\citet{xu09};
  40:~\citet{xu11}; 41:~\citet{xu13}; 42:~\citet{zhang09}}

\tablenotetext{\,}{Only a portion of the table is shown here to
  demonstrate its form and content.  A machine-readable version of the
  full table is available.}
\end{deluxetable}
%%%%%%%%%%%%%%%%%%%%%%%%%%%%%%%%%%%%%%%%%%%%%%%%%%

\clearpage
%%%%%%%%%%%%%%%%%%%%%%%%%%%%%%%%%%%%%%%%%%%%%%%%%%
\begin{deluxetable}{lcccccc}
\tabletypesize{\scriptsize}
\tablecaption{Galactic Distribution of WISE Catalog Sources$^{\rm a}$}
\tablewidth{0pt}
\tablehead{
\colhead{Quadrant} &
\colhead{Known} &
\colhead{Group} &
\colhead{Candidate} &
\colhead{Radio Quiet} &
\colhead{Total} &
\colhead{$^{12}{\rm CO\,}^{\rm b}$}
}
\startdata
First Quad. & 61 & 53 & 34 & 45 & 45 &    48\\ Second Quad. &  5 &  9 & 11 &  9 &  9 &     8\\ Third Quad. &  4 &  4 &  5 &  6 &  6 &  5\\ Fourth Quad. & 29 & 34 & 50 & 40 & 40 & 40\\ \hline $60\arcdeg$ of GC & 83 & 49 & 48 & 37 & 76 & 77\\ $30\arcdeg$ of GC & 49 & 24 & 30 & 30 & 45 & 52\\ $10\arcdeg$ of GC & 14 &  5 &  9 & 11 & 15 & 23\\ $\phn5\arcdeg$ of GC &  6 &  2 &  4 &  6 &  8 & 14\\ $\phn2\arcdeg$ of GC &  3 &  1 &  2 &  3 &  4 &  8\\ \hline $0\fdg10$ of $b=0\arcdeg$ & 23 & 21 & 20 & 24 & 22 & 11\\ $0\fdg25$ of $b=0\arcdeg$ & 49 & 39 & 41 & 45 & 44 & 26\\ $0\fdg50$ of $b=0\arcdeg$ & 71 & 65 & 62 & 66 & 65 & 44\\ \phn\phn$1\arcdeg$ of $b=0\arcdeg$ & 88 & 83 & 81 & 83 & 82 & 67\\

\enddata
\tablenotetext{a}{Entries list the percentage of WISE HII regions located within various Galactic zones.}
\tablenotetext{b}{From \citet{dame01}.}
\label{tab:quadrants}
\end{deluxetable}
%%%%%%%%%%%%%%%%%%%%%%%%%%%%%%%%%%%%%%%%%%%%%%%%%%

\end{document}